\documentclass[useAMS,usenatbib,fleqn]{mnras}
\usepackage{newtxtext}
\usepackage{mathptmx}
\usepackage{amssymb}
\usepackage{amsmath}
\usepackage{graphicx}
\usepackage[T1]{fontenc}
\DeclareRobustCommand{\VAN}[3]{#2}
\let\VANthebibliography\thebibliography
\def\thebibliography{\DeclareRobustCommand{\VAN}[3]{##3}\VANthebibliography}

\makeatletter
\DeclareRobustCommand{\Cpp}
{\valign{\vfil\hbox{##}\vfil\cr
   \textsf{C\kern-.1em}\cr
   $\hbox{\fontsize{\sf@size}{0}\textbf{+\kern-0.05em+}}$\cr}%
}
\makeatother

\title[Interference removal in pulsar observations]{Removal and replacement of interference in tied-array radio pulsar observations using the spectral kurtosis estimator}

\author[M.~Purver et al.]
{M.~Purver$^1$\thanks{\hspace{-0.1cm}E-mail: \href{mailto:mark.purver@alumni.manchester.ac.uk}{mark.purver@alumni.manchester.ac.uk}},
C.~G.~Bassa$^{1,2}$, I.~Cognard$^3$, G.~H.~Janssen$^{1,2}$, R.~Karuppusamy$^{1,4}$, \newauthor
M.~Kramer$^{1,4}$, K.~J.~Lee$^{1,4,5}$, K.~Liu$^{1,3,4}$, J.~W.~McKee$^6$, D.~Perrodin$^{1,7}$, S.~Sanidas$^{1,8}$, \newauthor
R.~Smits$^{1,2}$, B.~W.~Stappers$^1$ \\
$^1$Jodrell Bank Centre for Astrophysics, The University of Manchester, Manchester, M13 9PL, United Kingdom \\
$^2$ASTRON (Netherlands Institute for Radio Astronomy), Postbus 2, 7990 AA, Dwingeloo, the Netherlands \\
$^3$Station de Radioastronomie de Nan\c{c}ay, Observatoire de Paris, 18330 Nan\c{c}ay, France \\
$^4$Max Planck Institut f\"{u}r Radioastronomie, Auf dem H\"{u}gel 69, 53121 Bonn, Germany \\
$^5$Kavli Institute for Astronomy and Astrophysics, Peking University, Beijing 100871, P.~R.~China \\
$^6$Canadian Institute for Theoretical Astrophysics, University of Toronto, 60 St. George Street, Toronto, ON M5S 3H8, Canada \\
$^7$INAF -- Osservatorio Astronomico di Cagliari, Via della Scienza 5, 09047 Selargius (CA), Italy \\
$^8$Anton Pannekoek Institute for Astronomy, University of Amsterdam, Science Park 904, 1098 XH Amsterdam, the Netherlands
}

\let\leqslant=\leq

\date{Accepted 2021 November 22. Received 2021 November 16; in original form 2020 July 1}

\pubyear{2022}

\begin{document}
\label{firstpage}
\pagerange{\pageref{firstpage}--\pageref{lastpage}}
\maketitle

\begin{abstract}
We describe how to implement the spectral kurtosis method of interference removal (zapping) on a digitized signal of averaged power values. Spectral kurtosis is a hypothesis test, analogous to the t-test, with a null hypothesis that the amplitudes from which power is formed belong to a `good' distribution -- typically Gaussian with zero mean -- where power values are zapped if the hypothesis is rejected at a specified confidence level. We derive signal-to-noise ratios (SNRs) as a function of amount of zapping for folded radio pulsar observations consisting of a sum of signals from multiple telescopes in independent radio-frequency interference (RFI) environments, comparing four methods to compensate for lost data with coherent (tied-array) and incoherent summation. For coherently summed amplitudes, scaling amplitudes from non-zapped telescopes achieves a higher SNR than replacing zapped amplitudes with artificial noise. For incoherently summed power values, the highest SNR is given by scaling power from non-zapped telescopes to maintain a constant mean. We use spectral kurtosis to clean a tied-array radio pulsar observation by the Large European Array for Pulsars (LEAP): the signal from one telescope is zapped with time and frequency resolutions of $6.25\,\mathrm{ms}$ and $0.16\,\mathrm{MHz}$, removing interference along with $0.27\ \mathrm{per\ cent}$ of `good' data, giving an uncertainty of $0.25\,\mathrm{\mu s}$ in pulse time of arrival (TOA) for \mbox{PSR J1022$+$1001}. We use a single-telescope observation to demonstrate recovery of the pulse profile shape, with $0.6\ \mathrm{per\ cent}$ of data zapped and a reduction from $1.22$ to $0.70\,\mathrm{\mu s}$ in TOA uncertainty.
\end{abstract}

% Select between one and six entries from the list of approved keywords

\section{Introduction}
\label{intro}

Terrestrial sources of radio-frequency interference (RFI) hamper many astronomical radio observations, swamping the signal from outer space with intense and unwanted human-made radiation. Although it is usually impossible to subtract interference from useful information received at the same time and frequency, it is possible to remove or replace contaminated portions of data entirely, and so prevent them from affecting the integrated signal.

To remove interference, we wish to identify it using a fair and reliable method. One such method, a general signal processing technique termed `spectral kurtosis'\mbox{ \citep{dwyer83}}, was first applied to radio astronomy by \citet{ngl07}, subsequently refined by \citet{ng10a,ng10b} and applied again by \citet{gln10}. It is a statistical method that attempts to separate `Gaussian white noise' from everything else, on the assumption that the useful information resembles Gaussian white noise while the interference does not. It uses signal power to make a binary decision about whether to remove data, and it can be applied with fine resolution in time and frequency. We refer to the quantity used to make this decision as `the estimator'.

Pulsars are weak and rapidly varying astronomical radio sources that require observations with fine time resolution, so interference cannot easily be `averaged out' with integration across time, frequency or multiple telescopes. Spectral kurtosis has previously been used to remove interference from single-telescope pulsar observations from the Parkes, Lovell and Green Bank Telescopes \mbox{\citep{vs13,dlc+14,l16,kkl+19}}. We extend this approach to cover the case of a pulsar observation consisting of a sum of signals from an array of telescopes in independent interference environments, which results in a variable number of telescopes contributing to the observation at different times and frequencies. Since it is possible to compensate for contaminated data at one telescope using uncontaminated data from the rest of the array, we compare four methods of replacing the data that are lost from an array pulsar observation when interference is removed.

We apply one such method to a timing observation made as part of the Large European Array for Pulsars (LEAP) project (first shown in \mbox{\citealt[\S4.5]{bjk+16}}). This array comprises five widely separated radio telescopes, whose simultaneous pulsar observations can be summed coherently to produce a tied array of equivalent sensitivity to a single telescope of $195\,\mathrm{m}$ in diameter. Coherent summation is not always possible, so LEAP sometimes acts as an incoherently summed array equivalent to a $130$-$\mathrm{m}$ telescope. LEAP provides pulsar observations of greater sensitivity than any other existing steerable radio telescope, enabling precise timing of dynamical properties such as a pulsar's rotational period and proper motion. The project aims to use high-precision timing to detect gravitational waves, and interference removal is necessary to maintain its accuracy. Spectral kurtosis can be applied independently to each telescope's data, excising interference while sacrificing as little useful information as possible.

In Section~\ref{method} of this paper, we fully describe the implementation of the spectral kurtosis method of \citet{ng10b}. In Section~\ref{advs}, we explain advantages and disadvantages of this method of interference detection. In Section~\ref{effects}, we explore the effects of general interference removal on folded pulse profiles produced using observations from an array of telescopes, and look at how adverse effects can be mitigated while maximising signal-to-noise ratio (SNR). In Section~\ref{pulsar}, we summarise the application of spectral kurtosis to a LEAP observation. In Section~\ref{conc}, we conclude.

\section{Spectral kurtosis method of interference detection}
\label{method}

Each instance of the spectral kurtosis estimator is derived from a portion of the radio signal, and its value is a measure of the statistical properties of that portion. We use a hypothesis test to classify data as either `good' or `bad' based on the value of the estimator, where bad data are usually interference and good data are usually not. The null hypothesis is that data are good, because we know the values expected from good data. The classification provides evidence of whether a portion is {\it likely} to be good or bad in reality, but, since it examines a finite number of data, it cannot tell us whether the portion is definitely good or bad. We therefore remove portions that are {\it suspected} of being bad when they give estimator values that are outside specific limits, and we call this removal `zapping'.  We can control our level of suspicion, because the estimator allows us to define the fraction of good data that we are willing to zap mistakenly. In general, the more good data we are willing to lose, the more interference we will eliminate. If the amount of interference in an observation is small then we may zap more good data than bad, but the spoiling effect of even a small amount of dominant interference is usually sufficient to justify this.

In the following subsections, we explain how to calculate the estimator and its limits. We provide some example values for variables used in the calculations, as an aid to implementation of the method. Where numerical integration is required, we refer to routines within the GNU Scientific Library (\textsc{gsl}) for the \textsc{c} and \textsc{c++} programming languages. All variables used have real values, with imaginary quantities shown explicitly using the imaginary unit $i$.

\subsection{The radio signal}
\label{method:signal}

Although the estimator is a function of signal power, we begin with the amplitudes from which power is derived. The digitized signal initially consists of an evenly sampled time series of radio amplitudes, covering a fixed frequency bandwidth; we assume that the continuous signal has been limited to cover the same bandwidth before being sampled, so that its information is captured acccurately \citep{s1949}. Each time sample records the amplitude in either one or two polarization components, and each amplitude may be either real or complex (in the case of complex amplitudes, the imaginary part is simply a phase-shifted version of the real part in which each sinusoidal wave making up the signal has been shifted by $\frac{\pi}{2}$ radians, capturing the same information as a real signal of twice the sampling rate). The amplitudes are drawn from a set containing a fixed number of discrete values, stored using a corresponding number of `sampling bits', and we assume this set to be large enough to approximate a continuum of values for statistical purposes. Although the number of values does not change, the values themselves can be calibrated dynamically during an observation.

The frequency resolution of the signal can be improved, at the expense of time resolution, by performing discrete Fourier transforms (DFTs) on sequences of amplitudes; the more time samples are used in each DFT, the more the frequency resolution improves \citep[pp. 260--262]{bracewell00}. Although this is often referred to as `moving from the time domain to the frequency domain', the values that come out of a DFT are still amplitudes, with each amplitude representing the signal at one time and one frequency. DFTs of consecutive sequences of amplitudes can therefore be used as a time series with multiple frequency channels, where each DFT contributes one time point to all channels. The process simply exchanges time resolution for frequency resolution, and it is worth noting that the DFT of a single value gives the value itself, i.e. the time-domain signal can be thought of as being the frequency-domain signal with one channel. The channelised amplitudes are generally complex, representing the magnitude and phase of the signal at each time and frequency. Given the same signal and equipment, a complex time series consisting of $T$ samples and a real time series consisting of $2T$ samples produce almost the same amplitudes in channels $1$ to $T-1$ of their respective DFTs (where $T$ is a positive integer), as long as the frequencies in the signal are within the range of the channels. This similarity allows amplitudes from telescopes with real and complex sampling to be added together in the frequency domain \citep[\S4.1]{bjk+16}. There are some differences in the way that continuous signal frequencies are distributed into discrete bins, which can be mitigated by applying different weights to each bin; the differences are larger in the lowest- and highest-frequency channels, and these channels are often not used because they do not approximate the original signal well even if weights are used \citep[pp. 281, 288]{bracewell00}. The DFT amplitudes in channels $0$ and $T$ from the real time series are themselves always real, so they cannot generally be compared to the amplitude in channel $0$ from the complex time series, and the complex time series does not produce a channel $T$.

\subsection{The statistical distribution of the signal}
\label{method:distrib}
The identification of interference by the spectral kurtosis method is based on the probability density function (PDF) of a set of amplitudes, which gives the probabilities of a single amplitude taking any given value and which we refer to as the `distribution' of the set. The number of amplitudes in the set can be chosen at will, and the time and frequency ranges covered by the set are the time and frequency resolutions of zapping. 

Depending on the origin of the signal, the distribution can have different general forms (shapes), and other distinguishing characteristics such as different mean values. We can use spectral kurtosis to test whether each measured set of amplitudes is well described by a particular distribution. In a `good' signal of Gaussian white noise, the amplitudes (taking the real and imaginary parts as separate values if the signal is complex) are uncorrelated and have a Gaussian distribution with a mean of $0$ when collected over the time and frequency resolutions of zapping and across all polarizations; in a `bad' signal, the distribution of amplitudes when collected over these ranges is non-Gaussian, contains correlated amplitudes and/or does not have a mean of $0$. Although the portion of the signal from the pulsar might have a non-Gaussian distribution, it is typically too weak to substantially alter the total signal distribution, so a detectably bad signal is usually caused by interference. If a bad signal is caused by unpolarized interference, the distribution over time and frequency within a single polarization is bad, and two polarizations have the same distribution as one another; if a bad signal is caused by polarized interference, two polarization components are differently distributed. In all but pathological cases, the real and imaginary parts of a complex signal have the same distribution as one another.

\subsection{The estimator}
\label{method:estim}

The estimator is an unbiased estimate of the scaled variance divided by the square of the mean for a set of samples of summed power. It is referred to as `spectral' because it can be calculated separately for each frequency channel by using Fourier transforms, although it can also be calculated across the full bandwidth without leaving the time domain. It is called `kurtosis' because the variance of signal power involves the mathematical fourth power of amplitude values, but each element in the calculation is more generally the square of a sum of squares rather than a simple fourth power.

Each summed power value, $P_m$, is assembled as a sum of squares of amplitude values that are either real, $A_n$, or complex, $A_n+iB_n$:
\begin{equation}
\label{eqn:power_real}
P_m\ =\ \sum_{n=1}^{2N}A_n^2
\end{equation}
or
\begin{equation}
\label{eqn:power_complex}
P_m\ =\ \sum_{n=1}^N\left (A_n^2+B_n^2\right )
\end{equation}
In general, a set of power values has a non-zero mean, referred to as a `baseline'. The baseline is usually subtracted from the values at some stage, but subtraction should not be done prior to calculation of the estimator. Our variable $N$ is equivalent to the product $Nd$ in \citet{ng10b}. $N$ counts real and imaginary numbers and ranges in time, frequency and polarization without distinction, e.g. $N=2$ could result from a sum of complex amplitudes at one time, one frequency and two polarizations or from a sum of real amplitudes at two times, two frequencies and one polarization. $2N$ is thus the total number of squared values that are summed to form each power value. For real amplitudes, $2N$ is a positive integer; for complex amplitudes, $N$ is a positive integer. The time and frequency ranges are usually contiguous and evenly sampled, while the polarization range usually covers two orthogonal modes, although statistically these conditions are not necessary. When forming each power value, the use of one time and one frequency makes the estimator most sensitive to bad data \citep{ng10b}, and the use of two polarizations allows it to be equally sensitive to polarized interference coming from different directions. The use of more than one time or frequency may be made in order to save data storage space or processing time, and power values can be added together to accomplish this without needing to know their constituent amplitudes. Spectral kurtosis can be extended to cases in which good data have a non-Gaussian amplitude distribution, resulting in non-integer values of $2N$ and a modification of equations~(\ref{eqn:power_real}) and (\ref{eqn:power_complex}), and this generalization has been used for 2-bit data that do not approximate a continuous signal \mbox{\citep{nkp19}}. Non-integer values of $2N$ could also be used if the amplitudes making up each summed power sample were distributed with different variances (e.g. if two polarization channels had different gain levels) -- but we do not make use of this, preferring calibration to equalize variances prior to interference detection \citep[\S4.3]{bjk+16}.

If formed from Gaussian-distributed amplitudes, the PDF of a set of $M$ values of $P_m$ is a gamma distribution. This gamma power distribution has the useful property that its variance and the square of its mean both scale linearly with the square of the variance of the Gaussian amplitude distribution, as long as the mean of the amplitude distribution is $0$. We can therefore create a quantity that is independent of the amplitude variance of good data, which is the key motivation for defining the estimator \citep[as in][]{ng10b} as
\begin{equation}
\label{eqn:shat}
\hat S\,=\,\frac{\left (MN+1\right )V}{\left (M-1\right )\mu^2}\,=\,\frac{MN+1}{M-1}\left (\frac{M\sum_{m=1}^MP_m^2}{\left (\sum_{m=1}^MP_m\right )^2} - 1\right ),
\end{equation}
where $\mu$ is the mean of the set of summed power values, $V$ is its variance and $\frac{MN+1}{M-1}$ is a scaling and unbiasing factor (note that factors involving $M$ and $N$ can be brought inside the sums in equations~(\ref{eqn:power_real}), (\ref{eqn:power_complex}) and (\ref{eqn:shat}) to avoid the use of excessively large or small numbers during the calculations). The variance is scaled by the number of amplitudes contributing to each power value, so the factor would be $N$ if it were not for the additional need to correct bias in an estimate of $\frac{V}{\mu^2}$ derived from a finite set. $M$ is the total number of summed power samples that contribute to $\hat S$, and is therefore an integer; since variance is only meaningful for a set containing more than one value, we have the condition that $M\geq 2$. The set of $M$ values usually covers a contiguous block of time and frequency: larger values of $M$ coarsen the time and/or frequency resolutions of zapping but make the estimator more sensitive to bad data over the ranges of those resolutions \citep{ng10a}, with \citet{ngl07} advising that $M\geq 37$ is required to zap monochromatic interference. $\hat S$ represents an individual instance of the estimator, so we use $S$ as the range of values that $\hat S$ can take; since the power values used to calculate $\hat S$ are always real, we have the condition that $S\geq 0$.

The underlying probability distribution of $S$ can be revealed by making many measurements of $\hat S$. But the distribution for good data can also be approximated analytically, allowing it to be calculated more efficiently. The distribution depends on $M$ and $N$, but, for Gaussian-distributed amplitudes with a mean of $0$, it does not depend on the variance of those amplitudes. In other words, the estimator behaves in the same way for Gaussian amplitudes (good data) of any `loudness', and can thus be used to distinguish them from most non-Gaussian amplitudes (bad data). The estimator shares this property with the t-statistic \citep{s1908}, and in fact a t-test could be used to classify good and bad data using amplitudes instead of power values. We have not undertaken an interference detection comparison between the estimator and the t-statistic, but have employed the estimator because it can be used on either averaged power values or amplitudes and because it can apply a consistent test to polarized signals regardless of the angle between the radio source and the receiver plane.

For good data, $S$ has a mean of $1$ for all allowed values of $M$ and $N$; if we are to decide which amplitudes to accept as good without deriving the distribution of $S$ empirically, we must calculate the distribution's shape as best we can by computing some of its higher moments as well.

\subsection{The probability distribution of the estimator}
\label{method:prob}

In order to understand what values we expect the estimator to take when the null hypothesis is true, we determine the approximate cumulative distribution function (CDF), $P(S)$, that is produced by good $\hat S$ values. The CDF gives the fraction of good $\hat S$ values that are expected to fall at or below any level $S$, and its shape depends only on $M$ and $N$. 

A CDF is the integral of a PDF, $p(S)$, defined by
\begin{equation}
\label{eqn:cdf}
P(S)\ =\ \int_{S_{min}}^S p(s)\,\mathrm{d}s\ =\ 1-\int_S^{S_{max}} p(s)\,\mathrm{d}s,
\end{equation}
where $s$ is simply a variable of integration and where the PDF and CDF are defined in the range $S_{min}\leq S\leq S_{max}$ (so $P(S_{min})=0$ and $P(S_{max})=1$). \citet{gln10} found that the required CDF is complicated to calculate, because it is the integral of a skewed PDF. \citet{ng10a,ng10b} showed that the CDF can be well approximated in most cases by numerically integrating a PDF called a Pearson distribution, which is defined by four parameters (given by equation~(9) of \citealt{ng10b}) and allows up to the first four of its statistical moments to be matched to those of the true PDF. The first parameter is the mean or first raw moment, which we set equal to $1$. The second parameter is the variance or second central moment, and is given by
\begin{equation}
\label{eqn:mu2}
\mu_2\ =\ \frac{2M^2N\left (N+1\right )}{\left (M-1\right )\left (MN+2\right )\left( MN+3\right )}.
\end{equation}
Depending on the values of $M$ and $N$, the PDF required may be a Pearson distribution of Type I, Type IV or Type VI \citep{p1895,p1901}. To find out which Type, we use two parameters related to the third and fourth central moments (and therefore to skewness and kurtosis),
\begin{equation}
\label{eqn:beta1}
\beta_1=\,\frac{8\left (MN+2\right )\left (MN+3\right )\big(MN\left (N+4\right )-5N-2\big)^2}{\left (M-1\right )\left (MN+4\right )^2\left (MN+5\right )^2N\left (N+1\right )}
\end{equation}
and
\begin{equation}
\label{eqn:beta2}
\begin{split}
\beta_2 = & \,\frac{3\left (MN+2\right )\left (MN+3\right )}{\left (M-1\right )\left (MN+4\right )\left (MN+5\right )\left (MN+6\right )\left (MN+7\right )}\phantom{\Bigg)} \\
& \times\Big(M^3N^3\left (N+1\right )+M^2N^2\big(3N^2+68N+125\big) \\
& \phantom{\times|}-MN\big(93N^2+245N+32\big)+12\big(7N^2+4N+2\big)\Big) \\
& \times\phantom{\,}\frac{1}{N\left (N+1\right )},
\end{split}
\end{equation}
to define:
\begin{equation}
\label{eqn:kappa}
\kappa\ =\ \frac{\beta_1\left (\beta_2+3\right )^2}{4\left (4\beta_2-3\beta_1\right )\left (2\beta_2-3\beta_1-6\right )}
\end{equation}
(for any PDF, the quantities $\mu_2$ and $\beta_1$ are non-negative and $\beta_2\geq \beta_1+1$). We use Type I if $\kappa\leq 0$, Type IV if $0<\kappa<1$ and Type VI if $\kappa>1$ (special cases of Type V if $\kappa=1$ and Type III if $\kappa=\infty$ do not arise for the allowed values of $M$ and $N$). Fig.~1 of \citet{ng10b} shows which types correspond to different values of $M$ and $N$, demonstrating that Type IV is likely to be used if $N\leq 13.5$, Type VI if $N\geq 14$ and Type I only if $M\leq 9$. When calculating the CDF for all three Types, we use an additional parameter (which is positive for the allowed values of $M$ and $N$) giving the ratio of the third and second central moments:
\begin{equation}
\label{eqn:alpha1}
\alpha_1\ =\ \sqrt{\mu_2\beta_1}\ =\ \frac{4M\big(MN\left (N+4\right )-5N-2\big)}{\left (M-1\right )\left (MN+4\right )\left (MN+5\right )}.
\end{equation}

For each Type, we calculate location-scale transformations of the PDF and the CDF, $p'(S')$ and $P'(S')$, where
\begin{equation}
\label{eqn:P}
P'(S')\ =\ \int_{S'_{min}}^{S'} p'(s)\,\mathrm{d}s\ =\ P(S)
\end{equation}
when
\begin{equation}
\label{eqn:S}
S\ =\ aS'+\lambda
\end{equation}
and where the values of $a$ and $\lambda$ are calculated differently for each Type ($a$ is positive for any PDF). $S'$ is used because $p'(S')$ is a simpler form of a Pearson distribution than $p(S)$. The transformation gives $p(S)$ the required moments: for example, the mean of $p'(S')$ should always be found to be $\frac{1-\lambda}{a}$, as this ensures that $p(S)$ has a mean of $1$. In the following paragraphs, we explain how to calculate $P'(S')$, $a$ and $\lambda$ for each Type (note that all square roots refer to the non-negative value).

\subsubsection{The CDF using Type I}
\label{method:prob:i}
Type I corresponds to $\kappa\leq 0$, and applies only in some cases for which $M\leq 9$ (as long as $N\geq 0.5$, which covers all of its allowed values when calculating the estimator distribution caused by Gaussian white noise). Such small values of $M$ give a PDF with a large variance, making the estimator so insensitive to interference that it might not be considered useful \citep{ng10a}. Additionally, we have found that Type I does not provide a good approximation to the true PDF of $S$ as generated using simulated random numbers, even though it does approximate the first four moments well. The problem is worst at small $S$, where the Type I PDF gives a substantial probability of $S<0$, despite the fact that $\hat S$ cannot be negative. These negative values indicate that a Pearson distribution is inadequate when $M$ is small, and that it is necessary to approximate more than four moments or to derive the PDF empirically in these cases. However, since we do not know the scope of applications for which spectral kurtosis will be employed, we include this part of the method for completeness.

Following a method equivalent to that of \citet[pp. 217--220]{kendall94}, we match the first four moments of the true PDF by using the parameters
\begin{equation}
\label{eqn:c0}
c_0\ =\ \mu_2\left (4\beta_2-3\beta_1\right ),
\end{equation}
\begin{equation}
\label{eqn:c1}
c_1\ =\ \alpha_1\left (\beta_2+3\right )
\end{equation}
and
\begin{equation}
\label{eqn:c2}
c_2\ =\ 6+3\beta_1-2\beta_2
\end{equation}
to define:
\begin{equation}
\label{eqn:c}
c\ =\ \sqrt{c_1^2+4c_0c_2}.
\end{equation}
We then use these with the further parameter
\begin{equation}
\label{eqn:c3}
c_3\ =\ 15+9\beta_1-7\beta_2
\end{equation}
to define:
\begin{equation}
\label{eqn:n1}
n_1\ =\ 2 + \frac{c_3}{c_2}\bigg(\frac{c_1}{c}-1\bigg)
\end{equation}
and
\begin{equation}
\label{eqn:n2}
n_2\ =\ 2 - \frac{c_3}{c_2}\bigg(\frac{c_1}{c}+1\bigg)
\end{equation}
(of these seven quantities, only $c_3$ can be negative for the allowed values of $M$ and $N$ in the Type I case). The transformed CDF can then be calculated as
\begin{equation}
\label{eqn:Pi}
P'(S')\ =\ \frac{\int_0^{S'}s^{n_1-1}\left (1-s\right )^{n_2-1}\,\mathrm{d}s}{\int_0^{1}s^{n_1-1}\left (1-s\right )^{n_2-1}\,\mathrm{d}s}\ =\ \frac{\mathrm{B}(S'; n_1, n_2)}{\mathrm{B}(n_1, n_2)},
\end{equation}
where $\mathrm{B}(S'; n_1, n_2)$ and $\mathrm{B}(n_1, n_2)$ are the incomplete and complete beta functions respectively. This can be computed as the normalised incomplete beta function using \textsc{gsl}, or it can be found using numerical integration \citep[pp. 9, 21]{bateman53} as
\begin{equation}
\label{eqn:Pi2}
\begin{split}
P'(S')\ = & \ \int_0^{S'}\exp\Big(\left(n_1-1\right)\ln s+\left(n_2-1\right)\ln\left (1-s\right ) \\
& \ \phantom{\exp\exp\Big(\Big(}-\ln \mathrm{B}(n_1,n_2)\Big)\,\mathrm{d}s
\end{split}
\end{equation}
with
\begin{equation}
\label{eqn:lnbeta}
\begin{split}
\ln \mathrm{B}(n_1, n_2)\ = & \ \int_0^{\infty}\left (\frac{\big(e^{-n_1s}+e^{-n_2s}-e^{-\left (n_1+n_2\right )s}-e^{-s}\big)}{s\big(1-e^{-s}\big)}\right . \\
& \;\,\phantom{\exp\ln\ }-\frac{e^{-s}}{s}\Bigg)\,\mathrm{d}s,
\end{split}
\end{equation}
both of which approaches avoid the use of very large numbers during the calculation. The transformed CDF is defined in the range $0\leq S'\leq 1$.

The transformation from $S'$ to $S$ can be made using equation~(\ref{eqn:S}) with
\begin{equation}
\label{eqn:ai}
a\ =\ \frac{c}{c_2}
\end{equation}
and
\begin{equation}
\lambda\ =\ 1 - \frac{1}{2c_2}\left (\frac{c_1c_3}{2c_2-c_3}+c\right ).
\label{eqn:li}
\end{equation}
The transformed PDF is a beta distribution, which has a mean of $\frac{1-\lambda}{a}=\frac{n_1}{n_1+n_2}$ as required. Example Type I values are: $M=3$, $N=4$, $a=53.67$, $\lambda=-1.699$.

\subsubsection{The CDF using Type IV}
\label{method:prob:iv}
Type IV corresponds to $0<\kappa<1$. It applies in all cases for which $M\geq 240$ and $N\leq 13.5$, and in some cases for which $14\leq M\leq 239$ and $N\leq 13.5$, so it is most commonly used when few amplitudes are used to make each power value. It approximates the true PDF of $S$ well for all values of $M$ and $N$ to which it applies, working best when $M\gtrsim 25$.

Following \citet{ng10a}, we match the first four moments of the true PDF by using the parameter
\begin{equation}
\label{eqn:r}
r\ =\ \frac{6\left (\beta_2-\beta_1-1\right )}{2\beta_2-3\beta_1-6}
\end{equation}
to define:
\begin{equation}
\label{eqn:u}
u\ =\ 16\left (r-1\right )-\beta_1\left (r-2\right )^2.
\end{equation}
We then use these to define:
\begin{equation}
\label{eqn:w}
w\ =\ r\left (r-2\right )\sqrt{\frac{\beta_1}{u}}
\end{equation}
(these three quantities are all positive for the allowed values of $M$ and $N$ in the Type IV case). The transformed CDF can then be calculated as:
\begin{equation}
\label{eqn:Piv}
\begin{split}
P'(S')\ = & \ \frac{2^r\left |\Gamma\left (\frac{r+2+iw}{2}\right )\right |^2}{\pi\Gamma(r+1)}\int_{-\infty}^{S'}\frac{\exp\left (w\arctan s\right )}{\left (s^2+1\right )^{\frac{r+2}{2}}}\,\mathrm{d}s \phantom{\frac{x}{\bigg|}} \\
= & \ \int_{-\infty}^{S'}\exp\bigg(w\arctan s - \frac{\left (r+2\right )\ln\left (s^2+1\right )}{2} \\
& \ \phantom{\int_{-\infty}^{S'}\exp\bigg(} + r\ln 2 + 2\ln\bigg|\Gamma\Big(\frac{r+2+iw}{2}\Big)\bigg| \\
& \ \phantom{\int_{-\infty}^{S'}\exp\bigg(} - \ln\pi - \ln\Gamma(r+1)\bigg)\,\mathrm{d}s,
\end{split}
\end{equation}
where $\Gamma$ denotes the gamma function, i.e. $\Gamma(r+1)=\int_0^{\infty}s^re^{-s}\,\mathrm{d}s$. The second form of equation~(\ref{eqn:Piv}) avoids the use of very large numbers during the calculation. Numerical integration is needed (unless potentially computationally expensive hypergeometric series are used), but the log-gamma functions can be computed using \textsc{gsl}, or they can be found using further numerical integration \citep[p. 21]{bateman53} as
\begin{equation}
\label{eqn:lngamma}
\ln\Gamma(r+1)\ =\ \int_0^{\infty}\left (\frac{e^{-\left (r+1\right )s}-e^{-s}}{s\left (1-e^{-s}\right )}+\frac{re^{-s}}{s}\right )\,\mathrm{d}s
\end{equation}
and
\begin{equation}
\label{eqn:lngamma2}
\begin{split}
\ln\bigg|\Gamma\Big(&\frac{r+2+iw}{2}\Big)\bigg|\ =\ \Re\bigg[\ln\Gamma\Big(\frac{r+2+iw}{2}\Big)\bigg] \\
& = \ \int_0^{\infty}\Bigg(\frac{\cos\left (\frac{ws}{2}\right )e^{-\frac{\left (r+2\right )s}{2}}-e^{-s}}{s\left (1-e^{-s}\right )}+\frac{re^{-s}}{2s}\Bigg)\,\mathrm{d}s.
\end{split}
\end{equation}
The transformed CDF is defined in the range $-\infty\leq S'\leq \infty$.

The transformation from $S'$ to $S$ can be made using equation~(\ref{eqn:S}) with
\begin{equation}
\label{eqn:aiv}
a\ =\ \frac{\sqrt{\mu_2u}}{4}
\end{equation}
and
\begin{equation}
\label{eqn:liv}
\lambda\ =\ 1 - \frac{\alpha_1\left (r-2\right )}{4}
\end{equation}
(there is a typographical error in the definition of $a$ in equation~(57) of \citealt{ng10a}, in which $6$ should read $16$). Example Type IV values are: $M=1000$, $N=2$, $a=0.5008$, $\lambda=0.5593$.

\subsubsection{The CDF using Type VI}
\label{method:prob:vi}
Type VI corresponds to $\kappa>1$. It applies in all cases for which $M\geq 3$ and $N\geq 14$, and in some cases for which $3\leq M\leq 239$ and $N\leq 13.5$, so it is most commonly used when many amplitudes are used to make each power value. Like Type I, Type VI cannot identify interference fairly when $M$ is small: it gives a substantial probability of $S<0$ in some cases for which $M\lesssim 12$, even though $\hat S$ cannot be negative. In the more useful cases for which $M\gtrsim 25$, however, simulations with random numbers showed that Type VI provides a good approximation to the true PDF of $S$.

Following \citet{ng10b}, we match the first three moments (not four, as in the other cases) of the true PDF by using the parameter
\begin{equation}
\label{eqn:h}
h\ =\ 4 + \sqrt{\beta_1\left (\frac{1}{\mu_2}+4\right )+16}
\end{equation}
to define:
\begin{equation}
\label{alpha}
\alpha\ =\ \frac{1}{\alpha_1}\left (\mu_2\left (h\left (\frac{1}{\alpha_1}\left (\frac{8\mu_2}{\alpha_1}-1\right )+1\right )+4\right )+1\right ) - 1
\end{equation}
and
\begin{equation}
\label{beta}
\beta\ =\ 3 + \frac{2h}{\beta_1}.
\end{equation}
(these three quantities are all greater than 8 for the allowed values of $M$ and $N$ in the Type VI case). The transformed CDF can then be calculated as
\begin{equation}
\label{eqn:Pvi}
P'(S') =\,\frac{\int_0^{S'}s^{\alpha-1}\left (1+s\right )^{-\left (\alpha+\beta\right )}\,\mathrm{d}s}{\int_0^{1}s^{\alpha-1}\left (1-s\right )^{\beta-1}\,\mathrm{d}s}\,=\,\frac{\mathrm{B}\big(\frac{S'}{1+S'}; \alpha, \beta\big)}{\mathrm{B}\big(\alpha, \beta\big)},
\end{equation}
where $\mathrm{B}\big(\frac{S'}{1+S'}; \alpha, \beta\big)$ and $\mathrm{B}\big(\alpha, \beta\big)$ are the incomplete and complete beta functions respectively. This can be computed as the normalised incomplete beta function using \textsc{gsl} (in which case the transformation can later be made directly from $\frac{S'}{1+S'}$ to $S$ without explicitly calculating $S'$), or it can be found using numerical integration as
\begin{equation}
\label{eqn:Piv2}
\begin{split}
P'(S')\ = & \ \int_0^{S'}\exp\Big(\left(\alpha-1\right)\ln s-\left(\alpha+\beta\right)\ln\left (1+s\right ) \\
& \ \phantom{\exp\exp\Big(\Big(}-\ln \mathrm{B}(\alpha, \beta)\Big)\,\mathrm{d}s
\end{split}
\end{equation}
with equation~(\ref{eqn:lnbeta}) for $\ln \mathrm{B}(\alpha, \beta)$,
both of which approaches avoid the use of very large numbers during the calculation. The transformed CDF is defined in the range $0\leq S'\leq \infty$.

The transformation from $S'$ to $S$ can be made using equation~(\ref{eqn:S}) with
\begin{equation}
\label{eqn:avi}
a\ =\ 1
\end{equation}
and
\begin{equation}
\label{eqn:lvi}
\lambda\ =\ 1 - \frac{\alpha}{\beta-1}.
\end{equation}
The transformed PDF is a beta-prime distribution, which has a mean of $\frac{1-\lambda}{a} = \frac{\alpha}{\beta-1}$ as required. Example Type VI values are: $M=600$, $N=16$, $a=1$, $\lambda=-0.3393$.

\subsection{The limits of the estimator}
\label{method:limits}

We now decide on the fraction of good data that we are willing to reject, $2f$, and use the transformed CDF to calculate lower and upper limits of $S$ such that a fraction $f$ of good $\hat S$ values are expected to fall below the lower limit, $S_L$, and a fraction $f$ are expected to fall above the upper limit, $S_U$. The range between the limits is a confidence interval: we reject any portion of data that produces an estimator value outside the range, because we suspect those data of being bad. We wish to choose the smallest value of $f$ that adequately removes interference, in order to keep as many good data as possible and avoid any substantial change to the overall amplitude distribution of an observation.

Since we are using $P'(S')$, we calculate transformed limits $S'_L$ and $S'_U$ that are related to $S_L$ and $S_U$ by equation~(\ref{eqn:S}). These must satisfy the condition:
\begin{equation}
\label{eqn:f}
P'(S'_L)\ =\ 1-P'(S'_U)\ =\ f,
\end{equation}
where $0\leq f\leq 1$. The transformed PDF has a mean of $\frac{1-\lambda}{a}$ and a variance of $\frac{\mu_2}{a^2}$, so initial guesses for $S'_L$ and $S'_U$ can be found using a rough Gaussian approximation for the PDF:
\begin{equation}
\label{eqn:SL}
S'_L\ \approx\  \frac{1-\lambda-\eta\sqrt{\mu_2}}{a}
\end{equation}
and
\begin{equation}
\label{eqn:SU}
S'_U\ \approx \ \frac{1-\lambda+\eta\sqrt{\mu_2}}{a},
\end{equation}
where $\eta$ is a positive number such that
\begin{equation}
\label{eqn:eta}
f\ =\ \frac{1}{2}-\frac{1}{\sqrt{\pi}}\int_0^{\frac{\eta}{\sqrt{2}}}\exp\left (-s^2\right )\,\mathrm{d}s\ =\ \frac{1-\mathrm{erf}\left (\frac{\eta}{\sqrt{2}}\right )}{2}
\end{equation}
and `erf' denotes the error function. $f$ can be calculated from $\eta$ using the error function in \textsc{gsl} or using numerical integration; alternatively, $\eta$ can be calculated from $f$ using the inverse cumulative Gaussian distribution function in \textsc{gsl}. The use of $\eta$ is not absolutely necessary, but it allows us to make reasonable initial guesses and to describe our non-Gaussian PDF using the familiar language of Gaussian distributions: if we choose $\eta=3$, for example, then we can refer to $S_L$ and $S_U$ as `three-sigma limits', meaning that they exclude the same fraction of good data from our PDF as limits that were three standard deviations from the mean would exclude from a Gaussian distribution (where standard deviation is the square root of variance).

After checking that the initial guess for a transformed limit falls within the range for which the relevant CDF is defined, we calculate the value of the CDF at that point. Since all CDFs increase monotonically, we make the guess a lower bound if the CDF falls below its target value or as an upper bound if the CDF falls above its target value. We then calculate the CDF at intervals in $S'$ (moving in one direction by, for example, $\frac{\sqrt{\mu_2}}{a}$ at a time) until the CDF crosses its target value, giving us the other bound for the transformed limit. After this, we bisect the upper and lower bounds iteratively, and at each iteration we make the bisection point a new upper or lower bound according to the value of the CDF at that point. Once the upper and lower bounds are sufficiently close together, we take their bisection point as the transformed limit, and finally convert this to a limit on $S$ using equation~(\ref{eqn:S}). We use our limits to zap portions of data that give $\hat S<S_L$ or $\hat S>S_U$.

For $\eta=3$ ($f=0.001350$), example values corresponding to those in Sections~\ref{method:prob:i}--\ref{method:prob:vi} are: $M=3$, $N=4$, $S_L=-1.492$, $S_U=7.417$ (Type I); $M=1000$, $N=2$, $S_L=0.8499$, $S_U=1.1818$ (Type IV); $M=600$, $N=16$, $S_L=0.8321$, $S_U=1.1901$ (Type VI).

\section{Advantages and disadvantages of the estimator}
\label{advs}

Spectral kurtosis is one of many methods designed to distinguish interference from useful data. While spectral kurtosis looks for non-Gaussianity in the distribution of amplitudes, most techniques flag outlying power values. `Median absolute deviation', for example, distinguishes any portion of data whose power is substantially different to the portions around it, using a median-based variance estimate that is robust to outliers \mbox{\citep{frid09}}. The pulsar processing software \textsc{psrchive} can tackle narrowband interference by automatically zapping data in frequency channels that stand out from the median channel power, and can also remove impulsive interference by zapping parts of an average pulse profile that deviate from the expected shape \mbox{\citep{vs12}}.

Other approaches to interference mitigation include: eliminating `cyclostationary' signals that have periodic statistical properties \mbox{\citep{ait2010}}; characterizing known sources of interference in a specific environment \mbox{\citep{czech18}}; and removing signals that appear simultaneously to both a telescope and an adjacent reference receiver that is not pointing at the astronomical source \mbox{\citep{briggs2000}}.

Since no method of interference removal is perfect, we examine some of the particular features of the estimator.

\subsection{Advantages of the estimator}
\label{advs:ad}

The estimator is statistically unbiased, allowing us to choose fairly the fraction of good data that will be rejected. It is also simple in its assumption that the distributions of interference can usually be distinguished from the distribution of useful data: this is an unashamedly frequentist approach that requires no prior knowledge of the interference distributions. An examination by eye shows that spectral kurtosis is successful in zapping many different kinds of interference, with the loss of only a small fraction of good data.

As shown in Section~\ref{pulsar}, the estimator can be used to zap interference with fine time and frequency resolutions (e.g. $6.25\,\mathrm{ms}$ and $0.16\,\mathrm{MHz}$), salvaging more useful data than other methods with coarser resolutions. Zapping can operate in either the time or frequency domain, allowing the time resolution to be improved at the expense of frequency resolution or vice versa. The ability of the estimator to detect interference generally improves as more power values are accumulated, since its variance decreases as $M$ increases \citep{ng10a}, but the best value of $M$ also depends on the distribution and timescale of the interference.

It is possible to zap different types of interference by using the estimator with different values of $M$ on the same data, either independently or jointly in a `multiscale' approach \citep{gln10}, the latter also allowing transient signals to be detected and classified using data with as few as 2 sampling bits \citep{n16,ng16,nkp19}. It is also possible to make the estimator more sensitive to randomly changing signals and less sensitive to smoothly varying ones by normalising the power at each frequency and time by the total power across the bandwidth at the same time, before using the normalised values to produce the estimator \citep{ngl07}.

The estimator can be effective on power values that have been averaged together, so it can be used retrospectively after data have been compressed in this way, although its performance deteriorates as more values are used in each average.

\subsection{Disadvantages of the estimator}
\label{advs:dis}

The frequentist approach that makes spectral kurtosis simple is also its fundamental limitation. Ideally, we would zap data using a Bayesian method in which we had prior knowledge of the amount and the distributions of the interference. Without these things, we cannot give an accurate probability that any particular portion of data is bad. But prior knowledge is not generally available to us, since the interference environment often changes on timescales shorter than the duration of our radio observations. We therefore choose to use spectral kurtosis based on our long-term experience and suspicions about interference, but must acknowledge that the technique will be more successful in some situations than others.

Inevitably, the estimator will sometimes cause us to zap useful data or fail to zap interference. This can happen in two ways. Firstly, data can be mislabelled as good or bad. We can control the fraction of good data that will be rejected (the type I error rate), but we cannot predetermine the fraction of bad data that will be accepted (the type II error rate), since it depends on the similarity between the distributions of estimator values from bad data and the distribution from good data. Correlated Gaussian noise might be labelled as good, for example, and \citet{ngl07} found that periodic interference with a duty cycle of $40$--$60\ \mathrm{per\ cent}$ can closely mimic a good data distribution. Secondly, useful data can have a `bad' distribution or interference can have a `good' distribution. The former case could occur if single pulses from a pulsar had an SNR close to or greater than $1$ in a single frequency channel; the latter case could result from interference that had uncorrelated Gaussian-distributed amplitudes whose variance remained approximately constant over the zapping timescale and bandwidth. Zapping the pulsar, in particular, could alter its apparent profile shape (see Section~\ref{effects:situations}). To combat these problems, we should use frequency channels that are too fine for single pulses to be seen above the observational noise, and we may need to employ other methods of interference removal as well as, or instead of, spectral kurtosis.

The statistical nature of spectral kurtosis makes it less suitable for data that have been averaged over many power values. As the number of power values averaged together ($N$) increases, the estimator loses its ability to distinguish between different gamma power distributions and therefore to detect certain types of interference \citep{ng10b}. The method works best on power values sampled near the Nyquist rate, requiring substantial data storage space and computational power when zapping.

\section{Effects of interference removal on pulse profiles}
\label{effects}

Regardless of the method used to identify interference, zapping alters our data. We look at situations where this may cause a problem for pulsar observations, and examine four methods of compensating for the lost data in array observations with telescopes in independent interference environments.

\subsection{Situations in which zapping may alter profile shape}
\label{effects:situations}

When making simultaneous radio observations using multiple telescopes in order to increase the SNR, the signals from each telescope may be summed coherently, using amplitudes with phase information, or incoherently, using power without phase information. When portions of the signal at each telescope are independently zapped prior to summation, the final signal has a variable number of contributing telescopes as a function of time and frequency. This can pose a problem for pulsar observations in particular, because of the importance of profile shape.

After summation over the available telescopes, a pulsar observation is typically summed as a function of the pulsar's rotational phase in a process called `folding'. This gives the `average pulse', or profile. Samples are summed incoherently in a number of phase `bins', giving a profile with phase and frequency resolution, to which time resolution is added by repeating the process \citep[pp. 165--166]{lk05}. Profiles give a higher SNR than individual pulses, which allows more accurate timing of the pulsar. Timing accuracy also relies on the profile shape remaining highly stable over time, and so every care is taken to avoid altering it instrumentally. Profile shape change could be caused by portions of a strong pulsar signal being zapped as interference, but, even when zapping is unrelated to pulsar emission, we must still understand its effect on the profile.

To quantify the profile change caused by zapping, we begin with the amplitude signal at a single telescope with no interference (the signal may be real or complex and in the time or frequency domain). We neglect variation between individual pulse measurement, which is caused by phenomena such as pulse jitter and interstellar scintillation \mbox{\citep[see e.g.][Fig.~1.1 and pp. 8, 92, 202]{lk05}}, and assume that all pulses are identical monochromatic waves whose magnitude can be described as a function of rotational phase only. The signal power consists of a source (pulsar) contribution and a gamma-distributed random noise contribution, where the noise comes from Gaussian-distributed random amplitudes. In a single time sample within a single frequency and polarization channel, before summation over $N$ or $2N$ values (see Section~\ref{method:estim}), the source power alone would have a mean of $h(\theta)$ and a variance of $0$ at a single phase value, $\theta$, while the random noise power alone would have a mean of $g$ and a variance of $g^2$ (where $h(\theta)$ and $g$ are both non-negative). Because the source and noise have already been added together as amplitudes, within which the two contributions were independent, the summed power has a mean of $N\big (h(\theta)+g\big )$ and a variance of $N\left (2h(\theta)g+g^2\right )$. When many samples of summed power are added together in folding, the central limit theorem dictates that the folded power follows an approximately Gaussian distribution. Assuming that $g$ does not change with time, the variance is approximately constant if $h(\theta)\ll g$, which is usually the case since $h(\theta)$ and $g$ come from individual power samples and pulsar radiation at the Earth is weak. However, we can see that there is some phase-dependence to the variance, which could cause measured profile shape to deviate from the true profile shape given by the function $h(\theta)$ if the noise is non-Gaussian. Furthermore, zapping would cause variations in the power level that might manifest as additional phase-dependent noise or as an inconstant power baseline, where the baseline without zapping has a constant value of $Ng$.

\subsection{Equalizing zapped data in array observations with independent zapping at each telescope}
\label{effects:equalize}

After summing a signal coherently over multiple telescopes, we may wish to return the resulting amplitudes to the time domain via an inverse Fourier transform. We can avoid artefacts from the process by minimizing the noise level changes that can be produced by zapping, which also produces a more constant power baseline in both coherently and incoherently summed signals. This minimization is achieved by equalizing the amplitude or power variances of summed samples that have different amounts of zapping, either by scaling up samples that have fewer contributing telescopes or by adding artificial noise to them. If we are not returning the signal to the time domain, we can instead equalize the power mean by subtracting different baseline values from summed samples that have different amounts of zapping. Below, we compare the typical SNR of the summed signal after using these three processes and after no equalization.

If we assume that a fraction, $q$, of samples are zapped independently at each of $L$ telescopes, then the number of telescopes contributing to each summed sample is drawn from a binomial probability distribution. When many samples are zapped independently and added together in folding, and $q$ is not very close to either $0$ or $1$, the central limit theorem again shows that the folded signal power follows an approximately Gaussian distribution. The zapping introduces a characteristic mean, $\mu(x)$, and variance, $\sigma^2(x)$, given by:
\begin{equation}
\label{eqn:mu_zap}
\mu(x)\ =\ \frac{1}{L^x}\sum_{l=1}^Ll^xb_l
\end{equation}
and
\begin{equation}
\label{eqn:sigma_zap}
\sigma^2(x)\ = \ \mu(2x) - \mu(x)^2,
\end{equation}
where
\begin{equation}
\label{eqn:binom}
b_l\ =\ \frac{L!\,q^{L-l}\left (1-q\right )^l}{l!\left (L-l\right )!}
\end{equation}
and where $0\leq q\leq 1$ and $x$ is a positive number. Since $b_l$ is the probability of $l$ telescopes contributing to a sample, where $l$ is an integer in the range $0\leq l\leq L$, we find that $\sum_{l=0}^Lb_l=1$. The characteristic mean and variance are normalised by $L$ so that $0\leq\mu(x)\leq 1$ and $0\leq\sigma^2(x)\leq 0.25$ (if no zapping occurs, we have $q=0$, $\mu(x)=1$ and $\sigma^2(x)=0$). The value of $x$ depends on the method of variance equalization used and whether the summation of telescopes is coherent or incoherent: where $l$ telescopes contribute to a summed sample, the mean source contribution after equalization is proportional to $l^x$. Four useful identities are:
\begin{equation}
\label{eqn:mu_1}
\mu(1)\ =\ 1-q,
\end{equation}
\begin{equation}
\label{eqn:sigma_sq_1}
\sigma^2(1)\ =\ \frac{q\left (1-q\right )}{L},
\end{equation}
\begin{equation}
\label{eqn:sigma_sq_2}
\begin{split}
\sigma^2(2)\ = & \ \frac{4q\left (1-q\right )^3}{L} + \frac{\left (10q-4\right )q\left (1-q\right )^2}{L^2} \\
& \,+ \frac{q\left (1-q\right )-6q^2\left (1-q\right )^2}{L^3}
\end{split}
\end{equation}
and
\begin{equation}
\label{eqn:mu_3}
\mu(3)\ =\ \left(1-q\right)^3 + \frac{3q\left (1-q\right )^2}{L} + \frac{\left (3q-2\right )\left (1-q\right )+2\left (1-q\right)^3}{L^2},
\end{equation}
with $\mu(2)$ and $\mu(4)$ found by placing these identities into equation~(\ref{eqn:sigma_zap}).

In the following paragraphs, we provide expressions for the mean and variance of the zapped, equalized and folded signal using multiple telescopes, employing these relations for the mean (expectation), $\mathrm{E}$, and variance, $\mathrm{Var}$, of a set of values, $\{X\}$:
\begin{equation}
\label{eqn:mean_add}
\mathrm{E}\{X\}\ =\ \sum_{l=0}^L\mathrm{E}\{X_l\}b_l
\end{equation}
and
\begin{equation}
\label{eqn:var_add}
\begin{split}
\mathrm{Var}\{X\}\ = & \ \sum_{l=0}^L\mathrm{E}\{X_l^2\}b_l - \big (\mathrm{E}\{X\}\big )^2  \\
= & \ \sum_{l=0}^L\Big (\mathrm{Var}\{X_l\}+\big (\mathrm{E}\{X_l\}\big )^2\Big )b_l - \big (\mathrm{E}\{X\}\big )^2,
\end{split}
\end{equation}
where $\{X\}$ is the union of $L$ subsets of values, $\{X_l\}$, and where the numbers of members in the subsets follow a binomial probability distribution. Here, $\{X\}$ represents a set of signal values after summation across multiple telescopes. The values may be the magnitudes of amplitudes, or they may be power values. $\{X_l\}$ represents the subset of these values that are produced using $l$ telescopes. The validity of the expressions has been verified using simulations of Gaussian-distributed random numbers representing signal amplitudes. We look at coherent and incoherent summation, and at four methods of dealing with zapping: variance equalization by scaling, variance equalization by addition of artificial noise, mean equalization and no equalization at all. Mean values represent the pulse profile, and additive terms in their equations that do not depend on $h(\theta)$ are baselines that can be subtracted after folding. Variance values represent profile noise, and we give precedence to the terms that are largest in a typical pulsar observation, with the sum of all terms below the first line of each variance equation becoming equal to $0$ if there is no zapping. The profile SNR, a measure of the quality of an observation, is given by:
\begin{equation}
\label{eqn:snr}
\mathrm{SNR}\ =\ \frac{\mu^\prime(\theta)}{\sigma(\theta)},
\end{equation}
where $\mu^\prime(\theta)$ is the baseline-subtracted mean of the folded pulse profile and $\sigma(\theta)$ is the positive square root of its variance. Both of these are functions of the pulsar's rotational phase, but the latter is only weakly dependent when $h(\theta)\ll g$. We assume that all interference has been zapped, that the pulsar signal itself is too weak to cause zapping, that the time series at all telescopes are perfectly aligned and that the signals have been scaled so that all telescopes record equal power variance. We also assume that the same fraction of samples are zapped at each telescope and that all telescopes record the same mean power. If these last two assumptions were not correct, the forms of the equations would still be valid, but $\mu(x)$ and $\sigma^2(x)$ would have to change into sums of all possible combinations of telescopes, $L^x$ would have to change into weighted sums of contributing telescopes and $h(\theta)$ and $g$ would have to change into weighted average values for the source and noise contributions in a single power sample at a single telescope.

\subsubsection{Equalization of coherent observations}
\label{effects:equalize:coherent}

With coherent summation, $\{X_l\}$ represents a set of magnitudes of amplitudes summed over $l$ telescopes. Power values are the squared magnitudes of these summed amplitudes. The set of power values has a measurable mean, $\mathrm{E}\{X_l^2\}$, which is $\big (l^2\overline h+lg\big )$ if the amplitudes are complex or $\big (l^2\overline h+lg\big )\big /2$ if they are real, where $\overline h$ is the time-average of $h(\theta)$ over all phase values. This quantity should be measured using a sufficient number of samples to make a stable estimate. Each summed amplitude is associated with its own value of $l$ after zapping, and we need to measure $\mathrm{E}\{X_l^2\}$ using all combinations of telescopes if $\overline h$ makes any substantial contribution, since the $\overline h$ and $g$ terms vary with different powers of $l$ and we cannot measure the two independently. If $g$ dominates, however, then we can calculate $\mathrm{E}\{X_l^2\}$ using a separate measurement from each telescope. Ideally, it should be measured and used separately in each frequency and polarization channel, but an average can be used if it is similar in each channel. We assume that $\overline h$ remains constant, as does $h(\theta)$ at each phase value, but any measurable variation in these parameters could be included in $g$, which would then acquire phase dependence, and the following equations for coherent and incoherent observations would remain valid. In principle, we could also replace $\overline h$ with $h(\theta)$ in these equations if we were able to measure the time average of $h(\theta)$ separately at each phase value, but this would require longer, folded variance measurements.

To equalize variances by scaling, we multiply each summed amplitude by $\sqrt{\left (L^2\overline h+Lg\right )\big /\left (l^2\overline h+lg\right )}$, except when $l=0$. We replace any sample that has been zapped at all telescopes with a random complex number in which the real and imaginary parts are drawn independently from a Gaussian distribution of mean $0$ and variance $\big (L^2\overline h+Lg\big )\big /2$ if the amplitudes are complex, or with a random real number drawn from the same distribution if the amplitudes are real. Summed power is then formed from the equalized amplitudes (as in equation~(\ref{eqn:power_real}) or (\ref{eqn:power_complex})), and $W$ consecutive power samples are summed into each profile phase bin. The summation continues over $F$ pulses, yielding a folded profile with a phase-dependent mean, $\mu_{C,S}(\theta)$, baseline-subtracted mean, $\mu^\prime_{C,S}(\theta)$, and variance, $\sigma^2_{C,S}(\theta)$, given by using $X^2$ in place of $X$ in equations~(\ref{eqn:mean_add}) and (\ref{eqn:var_add}):
\begin{equation}
\label{eqn:mean_scale_coherent}
\mu_{C,S}(\theta)\ = \ FWN\left (L^2\overline h+Lg\right )\sum_{l=0}^L\frac{\big (lh(\theta)+g\big )b_l}{l\overline h+g},
\end{equation}
\begin{equation}
\label{eqn:eff_mean_scale_coherent}
\mu^\prime_{C,S}(\theta)\ = \ FWN\left (L^2\overline h+Lg\right )\sum_{l=0}^L\frac{lh(\theta)b_l}{l\overline h+g},
\end{equation}
and
\begin{equation}
\label{eqn:var_scale_coherent}
\begin{split}
\sigma^2_{C,S}(\theta)\ = & \ FWN\left (Lg+L^2\overline h\right )^2\sum^L_{l=0}\frac{\left (g^2+2lh(\theta)g\right )b_l}{\left (g+l\overline h\right )^2} \\
& + F^zW^yN^2\left (Lg+L^2\overline h\right )^2\sum^L_{l=0}\frac{\big (g+lh(\theta)\big )^2b_l}{\left (g+l\overline h\right )^2} \\
& - F^zW^yN^2\left (Lg+L^2\overline h\right )^2\left (\sum^L_{l=0}\frac{\big (g+lh(\theta)\big )b_l}{g+l\overline h}\right )^2,
\end{split}
\end{equation}
where subscripts $C$ and $S$ indicate coherent summation and equalization by scaling respectively, and where $1\leq y\leq 2$ and $1\leq z\leq 2$. If $L\overline h<g$, as is usually the case, Taylor series allow the profile mean, baseline-subtracted mean and variance to be expressed as
\begin{equation}
\label{eqn:mean_scale_coherent_approx}
\begin{split}
\mu_{C,S} (\theta)\ = & \ FWNL^2\Big (\big (\mu(1)+T_0\big )h(\theta) + \big (1-\mu(1)-T_0\big )\overline h\Big ) \\
& + FWNLg,
\end{split}
\end{equation}
\begin{equation}
\label{eqn:eff_mean_scale_coherent_approx}
\mu^\prime_{C,S} (\theta)\ =\ FWNL^2\big (\mu(1)+T_0\big )h(\theta)
\end{equation}
and
\begin{equation}
\label{eqn:var_scale_coherent_approx}
\begin{split}
\sigma^2_{C,S}(\theta)= & FWN\Big (L^2g^2 + 2L^3\mu(1)h(\theta)g + 2L^3\big (1-\mu(1)\big )\overline hg\Big ) \\
& + 2FWNL^4\big (2\mu(1)-2\mu(2)-T_1\big )\overline hh(\theta) \\
& + FWNL^4\big (1+3\mu(2)-4\mu(1)+T_2\big )\overline h^2 \\
& + F^zW^yN^2L^4\sigma^2(1)\big (h(\theta)-\overline h\big )^2 \\
& + 2F^zW^yN^2L^4\big (\mu(1)-\mu(2)\big )\big (\overline hh(\theta)-\overline h^2\big ) \\
& - 2F^zW^yN^2L^4\big (\mu(1)T_0+T_0^2\big )\big (h(\theta)-\overline h\big )^2 \\
& + F^zW^yN^2L^4\big (2T_1-T_2\big )h(\theta)^2 \\
& - 2F^zW^yN^2L^4\big (T_0+T_1\big )\overline hh(\theta) \\
& + F^zW^yN^2L^4\big (2T_0+T_2\big )\overline h^2 \\
& - 2F^zW^yN^2L^3T_0\left (h(\theta)-\overline h\right )g,
\end{split}
\end{equation}
where
\begin{equation}
\label{eqn:big_sum_0}
T_0\ = \ \sum_{k=1}^\infty\frac{L^k\left (-\overline h\right )^k}{g^k}\big (\mu(k+1)-\mu(k)\big ),
\end{equation}
\begin{equation}
\label{eqn:big_sum_2}
\begin{split}
T_1\ =\ \sum_{k=1}^\infty\frac{L^k\left (-\overline h\right )^k}{g^k}\Big (&k\mu(k)-2\big (k+1\big )\mu(k+1) \\
& +\big (k+2\big )\mu(k+2)\Big )
\end{split}
\end{equation}
and
\begin{equation}
\label{eqn:big_sum_3}
\begin{split}
T_2\ =\ \sum_{k=1}^\infty\frac{L^k\left (-\overline h\right )^k}{g^k}\Big (&\big (k+1\big )\mu(k)-2\big (k+2\big )\mu(k+1) \\
& +\big (k+3\big )\mu(k+2)\Big ),
\end{split}
\end{equation}
and the terms in the sums become smaller as $k$ increases.

Since groups of $M$ summed power samples are zapped together, the values of $y$ and $z$ depend on whether groups of samples being folded into each profile bin are zapped independently, and their ranges are $1\leq y\leq 2$ and $1\leq z\leq 2$. If $M\gg W$, then $y\simeq 2$ (most consecutive samples entering a bin are zapped together). The value of $y$ decreases when $M$ decreases towards $W$, but with the `resonances' at which $y=2$ if zapping and binning are synchronised (e.g. if $M=W$ and the $M$ samples that are considered for zapping are the same set as the $W$ samples that are summed into a bin). When $M$ decreases below $W$, $y$ decreases until $y\simeq 1$ when $M\ll W$ (many consecutive groups of samples entering a bin are zapped independently), although this limit is reached very slowly unless $M$ is small. A baseband observation of a pulsar with a period of $10\,\mathrm{ms}$ using $0.16$-$\mathrm{MHz}$-wide frequency channels and $1024$ bins across the profile, with $N=2$ and $M=1000$, would give $y\simeq 2$, but a similar observation of a pulsar with a period of $10\,\mathrm{s}$ would give $y$ a lower value (though not close to $1$ unless $M$ were made smaller). If $M\leq WD$, then $z=1$, where $D$ is the number of bins across the pulse profile and so $WD$ is the number of summed power samples across a single pulse (all groups of samples entering a bin from different pulses are zapped independently). If $M>WD$ but $M\ll FWD$, then $z\simeq 1$, where $FWD$ is the number of samples across an entire profile (many groups of samples entering a bin from different pulses are zapped independently). When $M$ increases above $WD$ and towards $FWD$, $z$ increases until $z\simeq 2$ when $M\gg FWD$ (most groups of samples entering a bin from different pulses are zapped together). This last situation is undesirable as entire profiles would be zapped together, and $z\simeq 1$ is typical for a pulsar observation.

To equalize variances by adding artificial noise instead of scaling, we add a random number to each summed amplitude. If the amplitudes are complex, the random number is complex and has real and imaginary parts drawn independently from a Gaussian distribution of mean $0$ and variance $\big ((L^2-l^2)\overline h+(L-l)g\big )\big /2$; if the amplitudes are real, the random number is real and is drawn from the same distribution. A folded profile, formed from summed power as in the previous paragraph, then has a mean, $\mu_{C,A}(\theta)$, baseline-subtracted mean, $\mu^\prime_{C,A}(\theta)$, and variance, $\sigma^2_{C,A}(\theta)$, given by
\begin{equation}
\label{eqn:mean_add_coherent}
\mu_{C,A}(\theta)\ =\ FWN\Big (L^2\mu(2)h(\theta) + L^2\big (1-\mu(2)\big )\overline h + Lg\Big ),
\end{equation}
\begin{equation}
\label{eqn:eff_mean_add_coherent}
\mu^\prime_{C,A}(\theta)\ =\ FWNL^2\mu(2)h(\theta)
\end{equation}
and
\begin{equation}
\label{eqn:var_add_coherent}
\begin{split}
\sigma^2_{C,A}(\theta)= & FWN\Big (L^2g^2 + 2L^3\mu(2)h(\theta)g + 2L^3\big (1-\mu(2)\big )\overline hg\Big ) \\
& + 2FWNL^4\big (\mu(2)-\mu(4)\big )\overline hh(\theta) \\
& + FWNL^4\big (1+\mu(4)-2\mu(2)\big )\overline h^2 \\
& + F^zW^yN^2L^4\sigma^2(2)\left (h(\theta)-\overline h\right )^2,
\end{split}
\end{equation}
where subscript $A$ indicates equalization by addition of artificial noise.

If we do not wish to return our data to the time domain as coherently summed amplitudes, we may choose not to apply any variance equalization but still to apply mean equalization to the summed power samples in order to obtain a higher SNR. In this case, we add $N\left ((L^2-l^2)\overline h+(L-l)g\right )$ to each summed power sample, as this quantity involves the measurable mean of a set of summed power samples using $l$ telescopes. A folded profile, formed from summed power as in the previous paragraphs, then has a mean, $\mu_{C,M}(\theta)$, baseline-subtracted mean, $\mu^\prime_{C,M}(\theta)$, and variance, $\sigma^2_{C,M}(\theta)$, given by
\begin{equation}
\label{eqn:mean_coherent_subtract}
\mu_{C,M}(\theta)\ =\ FWN\Big (L^2\mu(2)h(\theta)+L^2\big (1-\mu(2)\big )\overline h+Lg\Big ),
\end{equation}
\begin{equation}
\label{eqn:eff_mean_coherent_subtract}
\mu^\prime_{C,M}(\theta)\ =\ FWNL^2\mu(2)h(\theta)
\end{equation}
and
\begin{equation}
\label{eqn:var_coherent_subtract}
\begin{split}
\sigma^2_{C,M}(\theta)\ = & \ FWN\left (L^2\mu(2)g^2+2L^3\mu(3)h(\theta)g\right ) \\
& + F^zW^yN^2L^4\sigma^2(2)\big (h(\theta)-\overline h\big )^2,
\end{split}
\end{equation}
where subscript $M$ indicates mean equalization.

If there has been very little zapping, we may choose not to use any equalization to mitigate its effects, at the cost of a slightly lower SNR. A folded profile, formed from summed power as in the previous paragraphs, then has a mean, $\mu_C(\theta)$, baseline-subtracted mean, $\mu^\prime_C(\theta)$, and variance, $\sigma^2_C(\theta)$, given by
\begin{equation}
\label{eqn:mean_coherent}
\mu_C(\theta)\ =\ FWN\big (L^2\mu(2)h(\theta)+L\mu(1)g\big ),
\end{equation}
\begin{equation}
\label{eqn:eff_mean_coherent}
\mu^\prime_C(\theta)\ =\ FWNL^2\mu(2)h(\theta)
\end{equation}
and
\begin{equation}
\label{eqn:var_coherent}
\begin{split}
\sigma^2_C(\theta)\ = & \ FWN\left (L^2\mu(2)g^2+2L^3\mu(3)h(\theta)g\right ) \\
& + F^zW^yN^2L^2\sigma^2(1)g^2 \\
& + 2F^zW^yN^2L^3\big (\mu(3)-\mu(1)\mu(2)\big )h(\theta)g \\
& + F^zW^yN^2L^4\sigma^2(2)h(\theta)^2.
\end{split}
\end{equation}

\subsubsection{Equalization of incoherent observations}
\label{effects:equalize:incoherent}

Incoherent summation generally produces a lower profile SNR than coherent summation, but may be necessary if amplitudes cannot be stored or if the alignment of time series between telescopes cannot be made accurate enough to guarantee coherence. With incoherent summation, $\{X_l\}$ represents a set of power values summed over $l$ telescopes. This set has a measurable mean, $\mathrm{E}\{X_l\}$, of $Nl\left (\overline h+g\right )$ and a measurable variance, $\mathrm{Var}\{X_l\}$, of $Nl\left (2\overline hg+g^2\right )$. These quantities should be measured using a sufficient number of samples to make stable estimates. Each summed power value is associated with its own value of $l$ after zapping, and we can calculate the means and variances of the summed power values for all combinations of telescopes using separate measurements from each telescope, since each quantity varies with a single power of $l$.

To equalize variances by scaling, we add \mbox{$N\big (L-l\big )\left (\overline h+g\right )$} to each summed power sample and then multiply the result by $\sqrt{L/l}$, except when $l=0$. We replace any sample that has been zapped at all telescopes with a random real number drawn from a gamma distribution of mean $NL\left (\overline h+g\right )$ (shifted from $NL\sqrt{2\overline hg+g^2}$ ) and variance $NL\left (2\overline hg+g^2\right )$. A folded profile, formed as in Section~\ref{effects:equalize:coherent}, then has a mean, $\mu_{I,S}(\theta)$, baseline-subtracted mean, $\mu^\prime_{I,S}(\theta)$, and variance, $\sigma^2_{I,S}(\theta)$, given by using equations~(\ref{eqn:mean_add}) and (\ref{eqn:var_add}):
\begin{equation}
\label{eqn:mean_scale_incoherent}
\mu_{I,S}(\theta)\ =\ FWNL\Big (\mu(\tfrac{1}{2})h(\theta)+\left (1-\mu(\tfrac{1}{2})\right )\overline h+g\Big ),
\end{equation}
\begin{equation}
\label{eqn:eff_mean_scale_incoherent}
\mu^\prime_{I,S}(\theta)\ =\ FWNL\mu(\tfrac{1}{2})h(\theta)
\end{equation}
and
\begin{equation}
\label{eqn:var_scale_incoherent}
\begin{split}
\sigma^2_{I,S}(\theta)\ = & \ FWNL\Big (g^2 + 2\big (1-q^L\big )h(\theta)g+2q^L\overline hg\Big ) \\
& + F^zW^yN^2L^2\sigma^2(\tfrac{1}{2})\left (h(\theta)-\overline h\right )^2,
\end{split}
\end{equation}
where subscript $I$ indicates incoherent summation.

To equalize variances by adding artificial noise instead of scaling, we add a random real number to each summed power sample. This number is drawn from a gamma distribution of mean $N\big (L-l\big )\left (\overline h+g\right )$ (shifted from \mbox{$N\big (L-l\big )\sqrt{2\overline hg+g^2}$ )} and variance $N\big (L-l\big )\left (2\overline hg+g^2\right )$. A folded profile, formed as in Section~\ref{effects:equalize:coherent}, then has a mean, $\mu_{I,A}(\theta)$, baseline-subtracted mean, $\mu^\prime_{I,A}(\theta)$, and variance, $\sigma^2_{I,A}(\theta)$, given by
\begin{equation}
\label{eqn:mean_add_incoherent}
\mu_{I,A}(\theta)\ =\ FWNL\Big (\mu(1)h(\theta)+\big (1-\mu(1)\big )\overline h+g\Big ),
\end{equation}
\begin{equation}
\label{eqn:eff_mean_add_incoherent}
\mu^\prime_{I,A}(\theta)\ =\ FWNL\mu(1)h(\theta)
\end{equation}
and
\begin{equation}
\label{eqn:var_add_incoherent}
\begin{split}
\sigma^2_{I,A}(\theta)\ = & \ FWNL\Big (g^2 + 2\mu(1)h(\theta)g+2\big (1-\mu(1)\big )\overline hg\Big ) \\
& + F^zW^yN^2L^2\sigma^2(1)\left (h(\theta)-\overline h\right )^2.
\end{split}
\end{equation}

We cannot return our incoherently summed power samples to meaningful amplitudes, and so we may choose not to apply any variance equalization but still to apply mean equalization in order to obtain a higher SNR. In this case, we add $N\big (L-l\big )\left (\overline h+g\right )$ to each summed power sample. A folded profile, formed as in Section~\ref{effects:equalize:coherent}, then has a mean, $\mu_{I,M}(\theta)$, baseline-subtracted mean, $\mu^\prime_{I,M}(\theta)$, and variance, $\sigma^2_{I,M}(\theta)$, given by
\begin{equation}
\label{eqn:mean_incoherent_subtract}
\mu_{I,M}(\theta)\ =\ FWNL\Big (\mu(1)h(\theta)+\big (1-\mu(1)\big )\overline h+g\Big ),
\end{equation}
\begin{equation}
\label{eqn:eff_mean_incoherent_subtract}
\mu^\prime_{I,M}(\theta)\ =\ FWNL\mu(1)h(\theta)
\end{equation}
and
\begin{equation}
\label{eqn:var_incoherent_subtract}
\begin{split}
\sigma^2_{I,M}(\theta)\ = & \ FWNL\mu(1)\left (g^2+2h(\theta)g\right ) \\
& + F^zW^yN^2L^2\sigma^2(1)\big (h(\theta)-\overline h\big )^2.
\end{split}
\end{equation}

If there has been very little zapping, we may choose not to use any equalization to mitigate its effects, at the cost of a slightly lower SNR. A folded profile, formed as in Section~\ref{effects:equalize:coherent}, then has a mean, $\mu_I(\theta)$, baseline-subtracted mean, $\mu^\prime_I(\theta)$, and variance, $\sigma^2_I(\theta)$, given by
\begin{equation}
\label{eqn:mean_incoherent}
\mu_I(\theta)\ =\ FWNL\mu(1)\big (h(\theta)+g\big ),
\end{equation}
\begin{equation}
\label{eqn:eff_mean_incoherent}
\mu^\prime_I(\theta)\ =\ FWNL\mu(1)h(\theta)
\end{equation}
and
\begin{equation}
\label{eqn:var_incoherent}
\begin{split}
\sigma^2_I(\theta)\ = & \ FWNL\mu(1)\left (g^2+2h(\theta)g\right ) \\
& + F^zW^yN^2L^2\sigma^2(1)\big (g+h(\theta)\big )^2.
\end{split}
\end{equation}

\subsection{Comparison of equalization methods for zapped data}
\label{effects:compare}

The choice of equalization method may depend on the fraction of data that are zapped, which can vary with time, frequency and telescope environment, and the choice may also depend on whether data are summed coherently or incoherently.  In most cases, mean equalization and variance equalization by scaling produce a higher profile SNR than variance equalization by addition of artificial noise and no equalization. This is because artificial noise introduces additional variance without increasing the mean, as does a failure to equalize the power baseline.

\begin{figure}
\begin{tabular}{ l }
\includegraphics[width=0.5\textwidth,clip=true,trim=4.1cm 0.1cm 0.55cm 1.52cm]{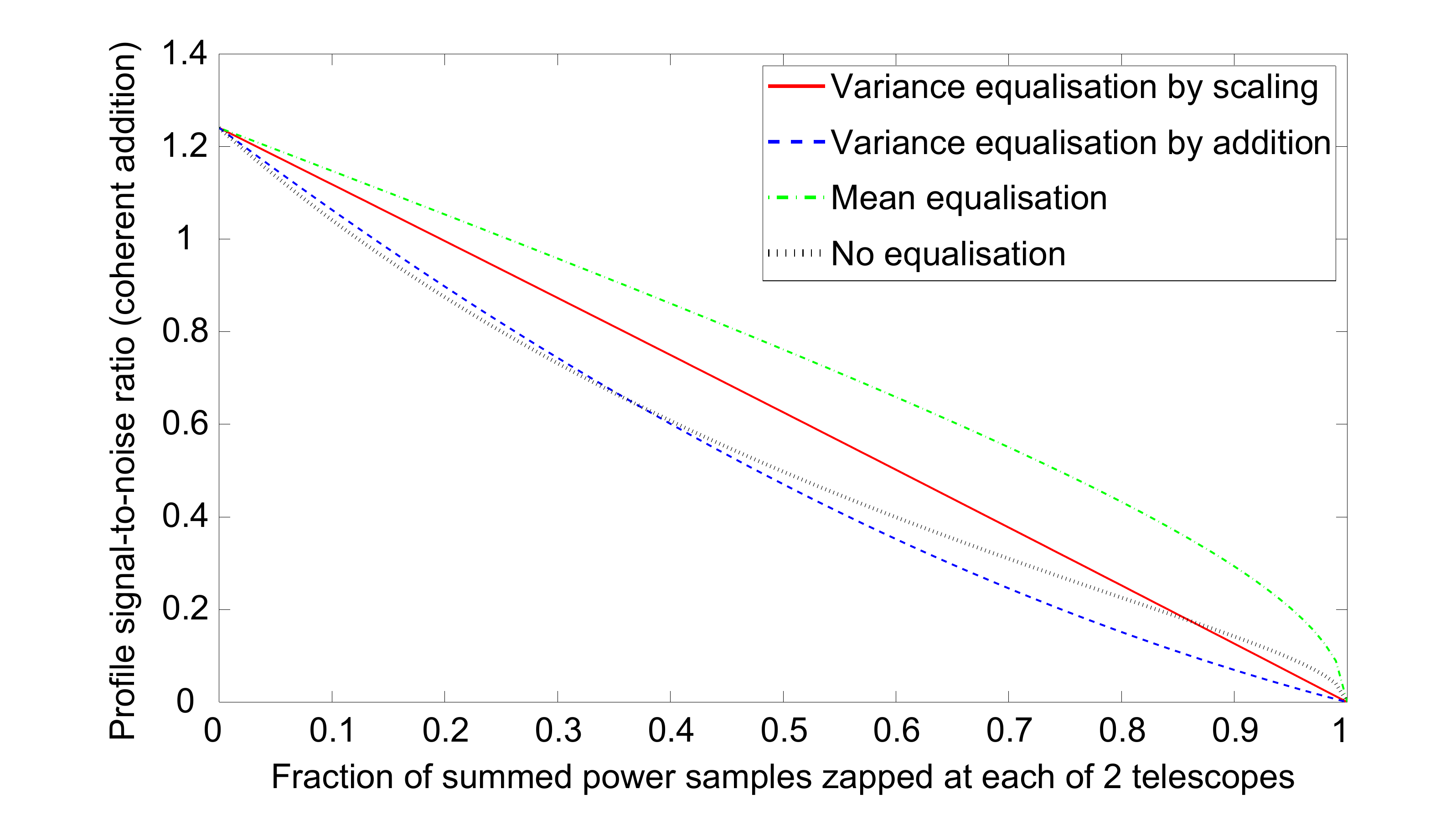} \\
\includegraphics[width=0.5\textwidth,clip=true,trim=3cm 0cm 1.5cm 0.92cm]{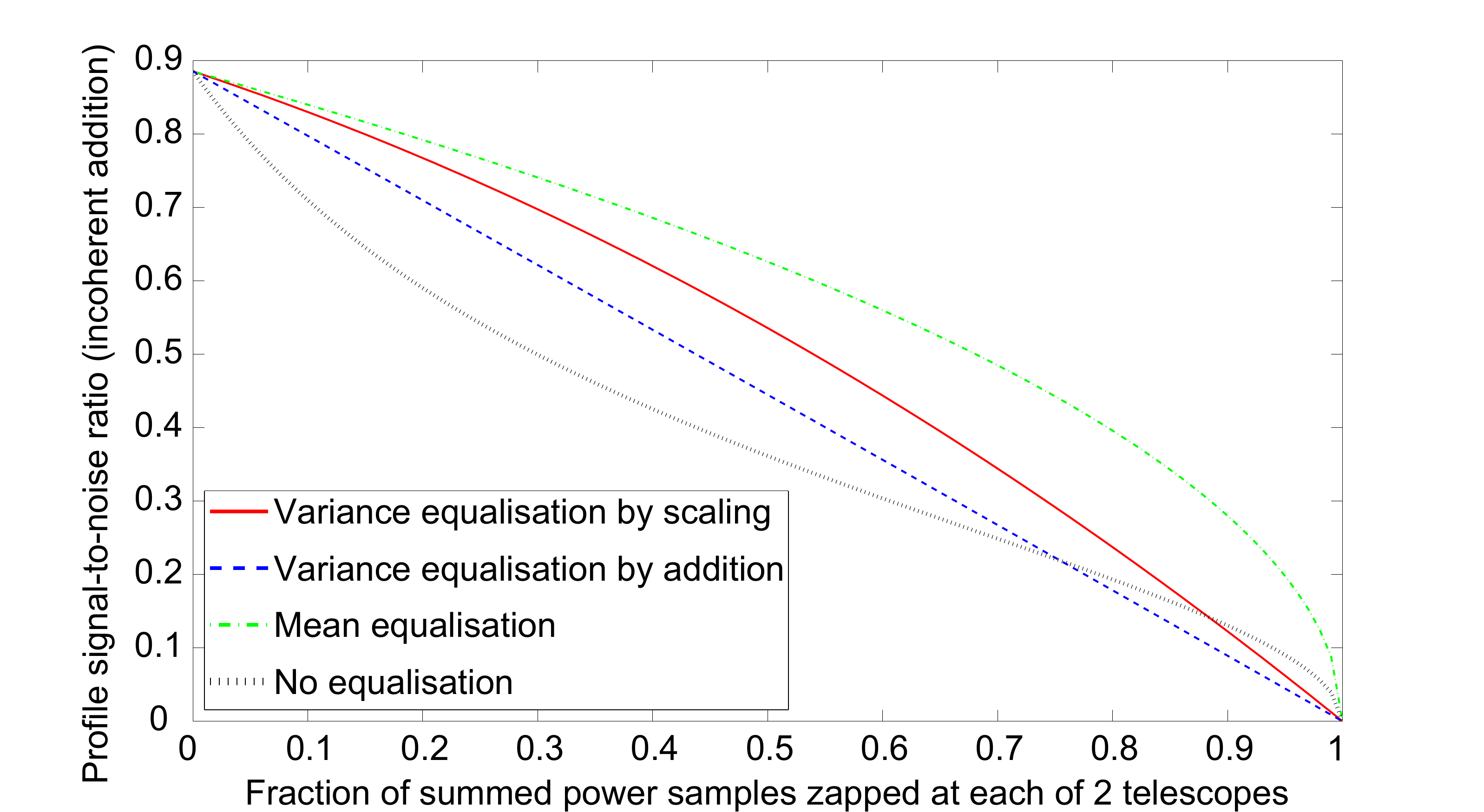}
\end{tabular}
\caption{Profile SNR as a function of the fraction of summed power samples zapped due to interference, using coherent (top) and incoherent (bottom) summation of signals from two identical telescopes in independent interference environments, with four different methods of equalization (see text for other parameter values).}
\label{fig:snr_2}
\end{figure}

\begin{figure}
\begin{tabular}{ l }
\includegraphics[width=0.466\textwidth,clip=true,trim=0cm 0cm 0cm 0cm]{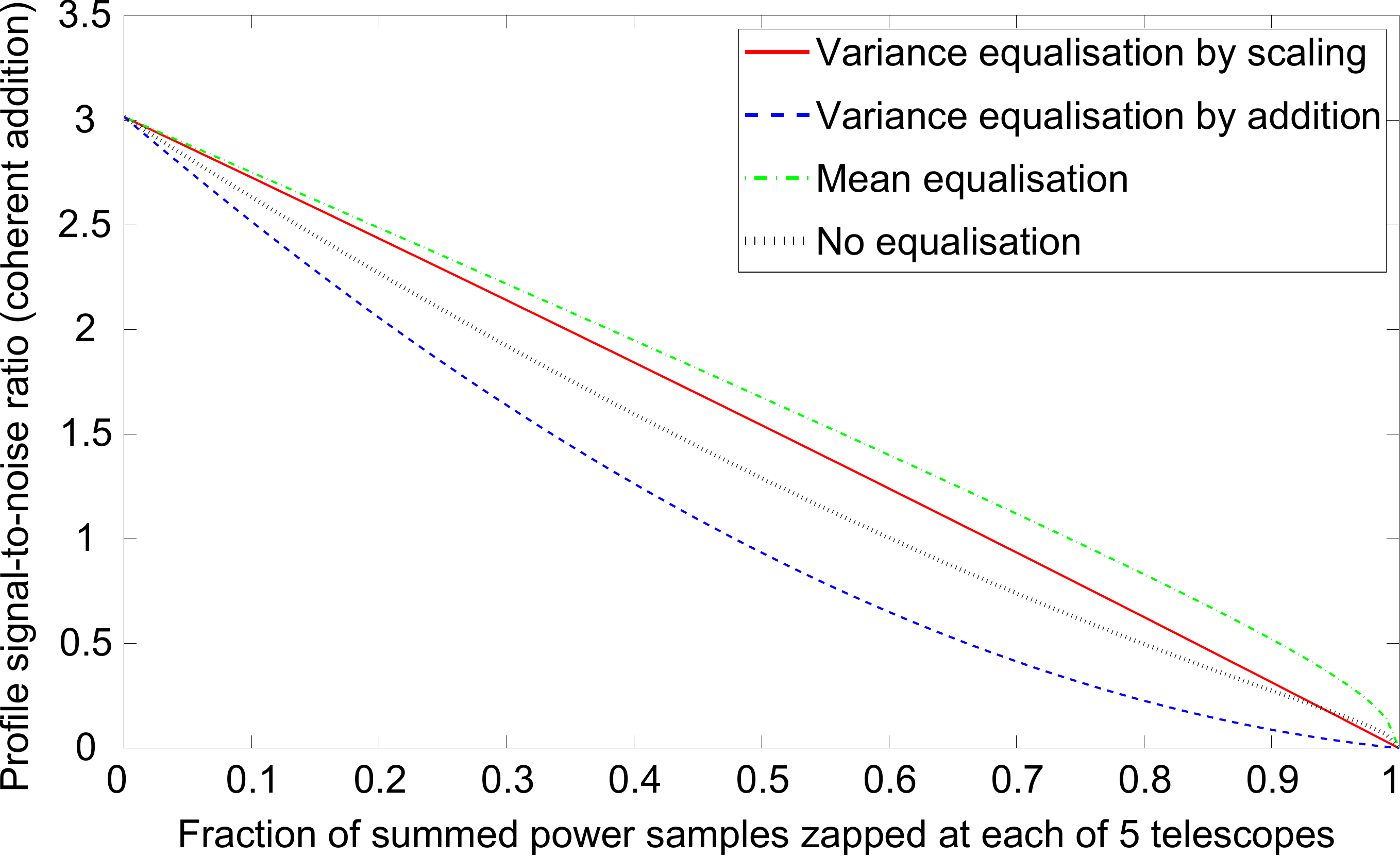} \\
\includegraphics[width=0.517\textwidth,clip=true,trim=3cm 0.08cm 0cm 0.49cm]{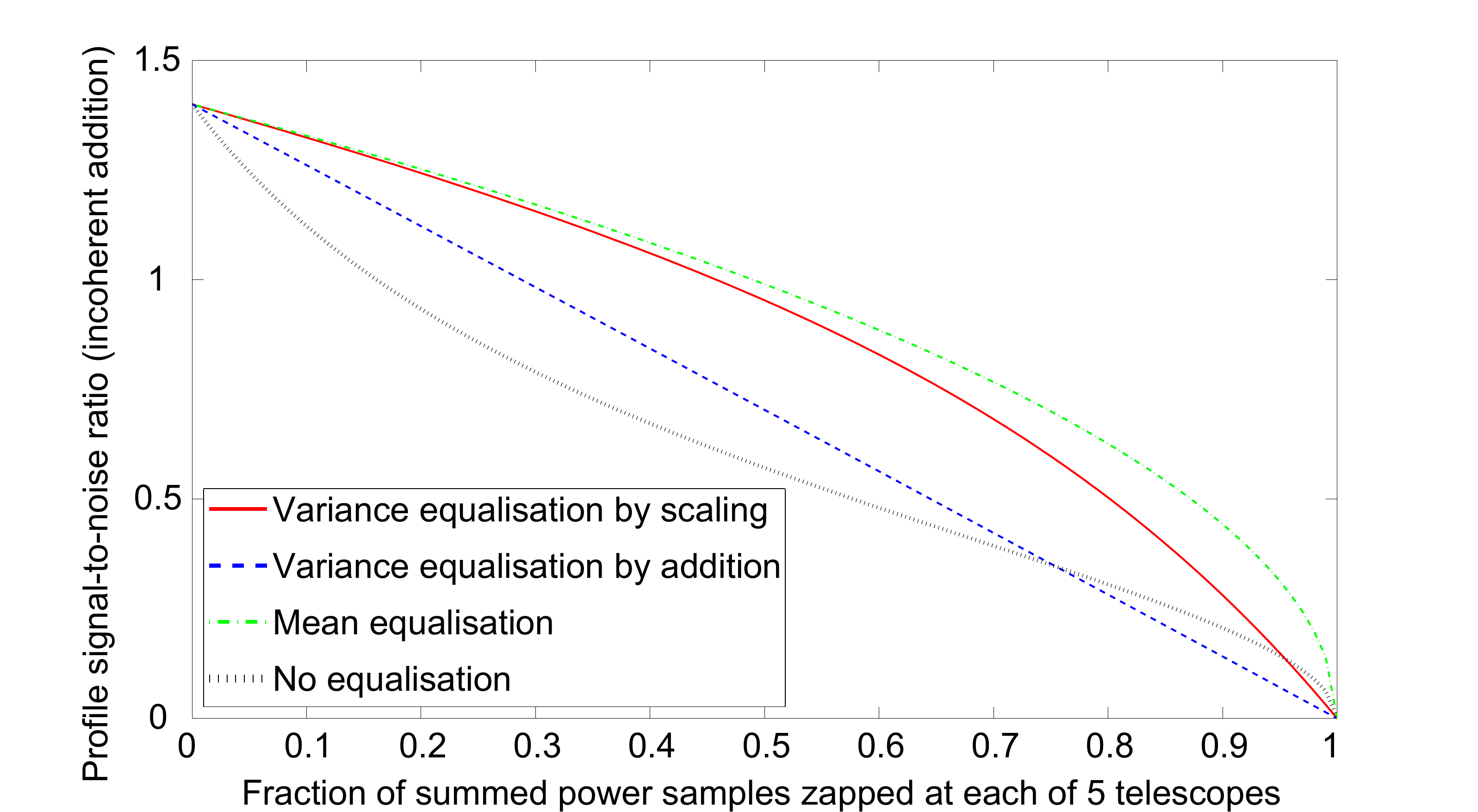}
\end{tabular}
\caption{Profile SNR as a function of the fraction of summed power samples zapped due to interference, using coherent (top) and incoherent (bottom) summation of signals from five identical telescopes in independent interference environments, with four different methods of equalization (see text for other parameter values).}
\label{fig:snr_5}
\end{figure}

\begin{figure}
\begin{tabular}{ l }
\includegraphics[width=0.515\textwidth,clip=true,trim=3cm 0cm 0cm 1.6cm]{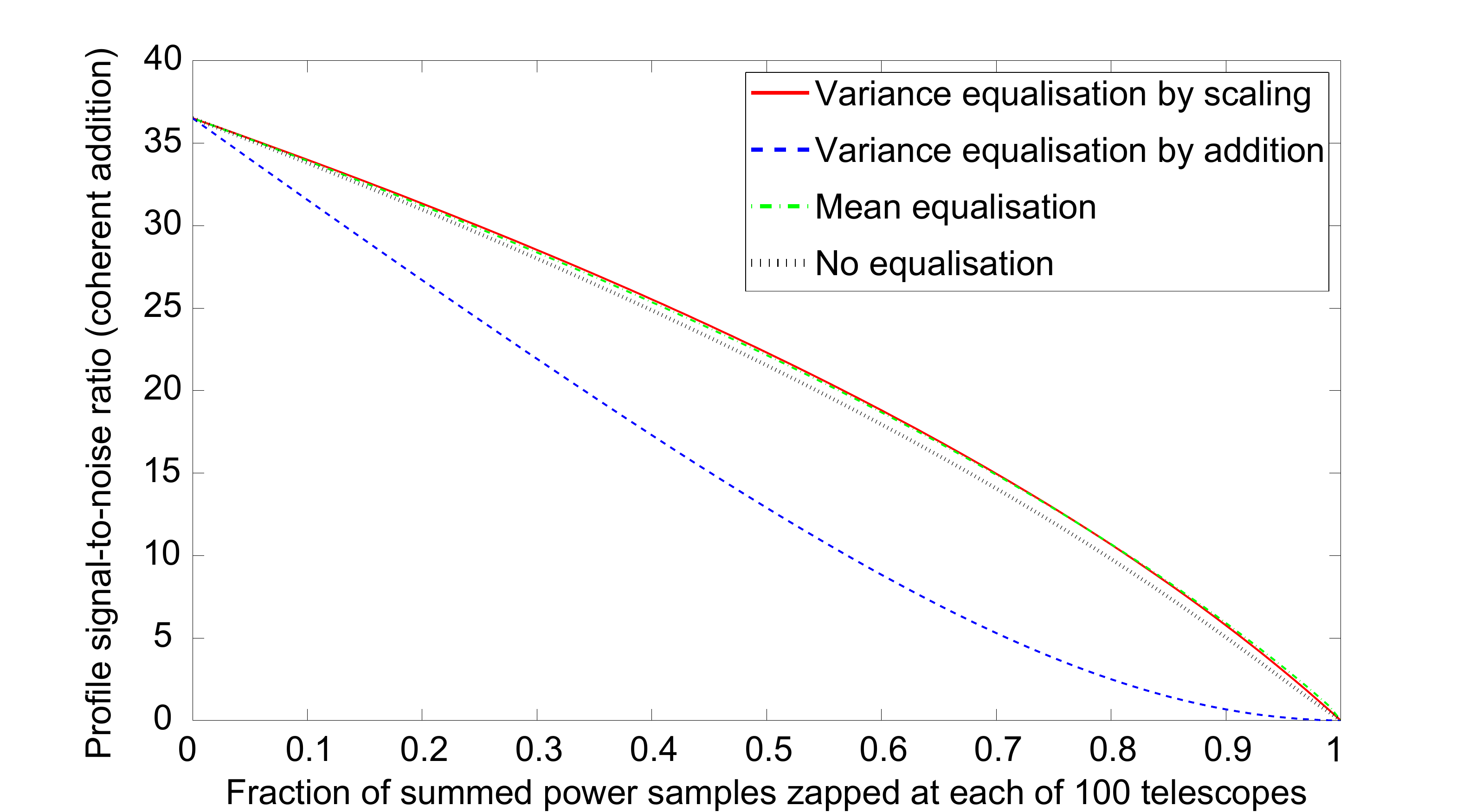} \\
\includegraphics[width=0.515\textwidth,clip=true,trim=3cm 0cm 0cm 0.2cm]{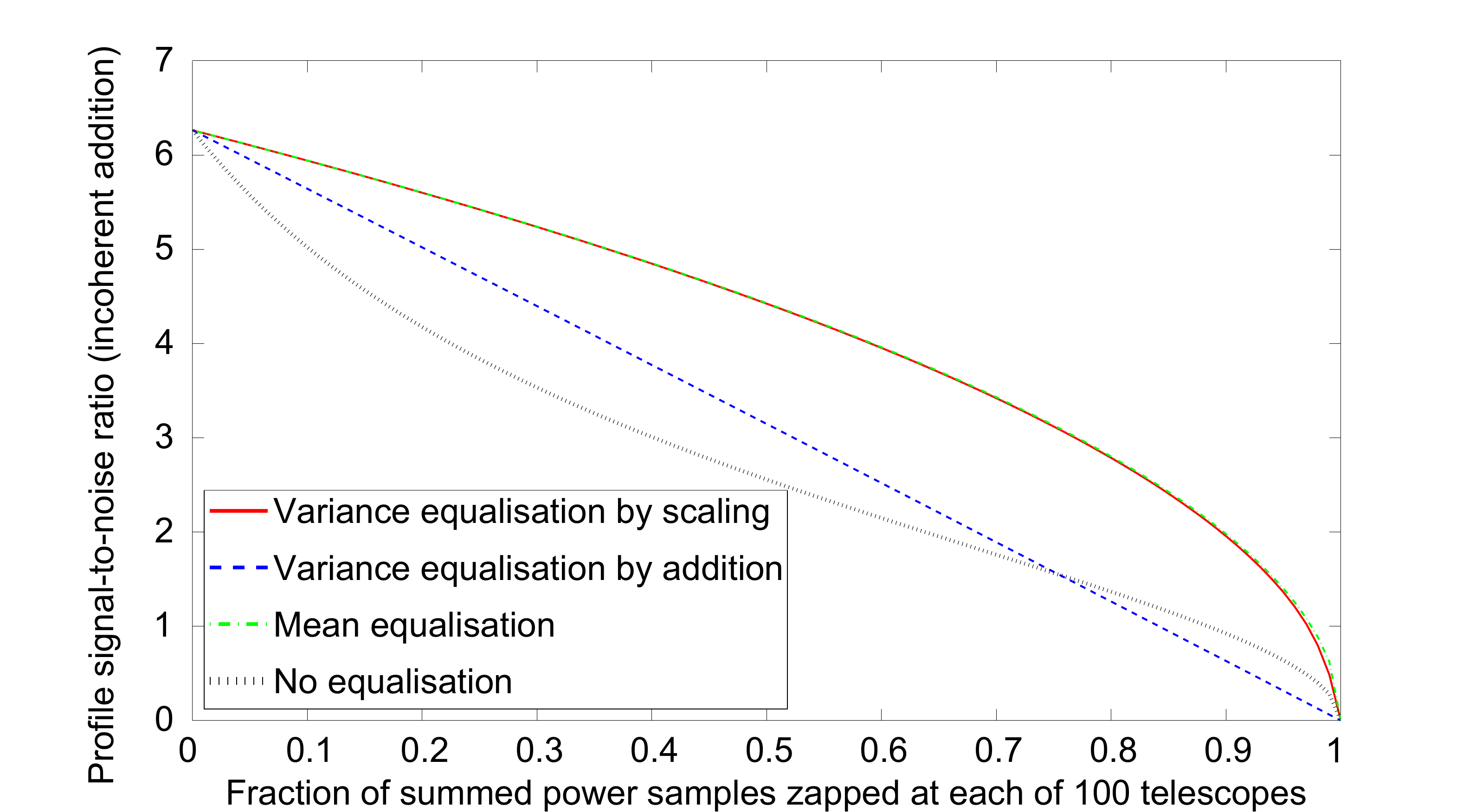}
\end{tabular}
\caption{Profile SNR as a function of the fraction of summed power samples zapped due to interference, using coherent (top) and incoherent (bottom) summation of signals from 100 identical telescopes in independent interference environments, with four different methods of equalization (see text for other parameter values).}
\label{fig:snr_100}
\end{figure}

We focus on the regime of $\overline h\ll h(\theta)\ll g$, which is typical of pulsar observations at phase values where the pulse is visible. The first inequality comes about because a normal pulsar gives no emission for the majority of its period, so the pulse itself is usually well above the average emission strength. The second inequality occurs because these are quantities in a single sample at a single telescope, in which noise usually dominates over source contribution (where noise does not dominate, coherent summation loses its SNR advantage over incoherent summation even if there is no zapping). Figs.~\ref{fig:snr_2}--\ref{fig:snr_100} use $h=0.01$, $\overline h=0.0001$, $g=1$, $N=2$, $W=2$, $y=2$, $F=1000$ and $z=1$ (representing an observation of a pulsar with a period of $10\,\mathrm{ms}$) and show the relationships between profile SNR ($\mu^\prime(\theta)/\sigma(\theta)$) and fraction of summed power samples zapped at each telescope ($q$) for 2, 5 and 100 identical telescopes in independent interference environments (this example also applies to identical groups of telescopes in which each group is zapped together and each group is in an independent interference environment; see \citealt{tjd+18} and \citealt{nh20} for calculations of the estimator using multiple receivers in the same environment). Coherent and incoherent summation are shown, with all four methods of equalization.

Mean equalization usually gives the highest SNR, although the difference between mean equalization and variance equalization by scaling decreases as more telescopes are added, and the second method is very slightly better for coherent summation of 100 telescopes with a small or moderate amount of zapping. The scaling method may be preferred for coherent summation in particular, because it avoids artefacts when returning amplitudes to the time domain (see Section~\ref{effects:equalize}), and the penalty in SNR is small unless there is a large amount of zapping. Mean equalization can fall behind if $W$ or $N$ increase sufficiently to cause the signal to make a substantial contribution to variance.

Variance equalization by addition of artificial noise lags some way behind the other active methods (except in the case of a single telescope, for which it is identical to variance equalization by scaling), but it gives a better SNR than the passive method of no equalization when summing two telescopes with a small amount of zapping, as well as preventing artefacts when returning coherently summed amplitudes to the time domain. With a larger number of telescopes, it only retains this SNR advantage for incoherent summation. It can be applied at each telescope without reference to the others, so it is computationally simpler than the scaling method and may be preferred if equalization needs to be done before the individual signals are combined -- for example, interference-free signals from widely spaced telescopes may be needed in order to synchronise their summation, after which the artificial noise could be replaced by a different equalization method.

The method of no equalization gives a relatively poor SNR for incoherent summation, but its performance for coherent summation improves as the number of telescopes increases. It is the simplest method, because it does not require replacement values to be computed, and may be preferred when there is very little zapping, or in the coherent summation of a large number of telescopes. However, its unequal power baseline can cause its SNR to decrease below that of all other methods when $W$ or $N$ increases, particularly for incoherent summation (e.g. when $W=2000$ and $y=1.5$, representing an observation of a pulsar with a period of $10\,\mathrm{s}$).

All four methods of equalization produce profile means that are linear functions of $h(\theta)$, so they do not make systematic changes to profile shape as long as the distribution of all profile noise is close to Gaussian. Even Gaussian profile noise with a phase-dependent variance produces Gaussian noise of constant variance in each complex-valued bin of a profile's DFT, which means that the standard method of frequency-domain pulsar timing should not produce unwanted correlations between timing residuals (often called `red noise') when using equalized signals \citep{tay92}. The main danger to profile shape, other than zapping of the pulsar contribution itself, is non-Gaussian noise. We rely on the accumulation of data into each profile phase bin to make the noise approximately Gaussian, in line with the central limit theorem (even though the equations for profile mean and variance above do not depend on the noise distribution). Substantial non-Gaussianity could arise if $q$ were very close to $1$, or if $q$ were very close to $0$ when $h\gtrsim g$. It could also occur if the duration and bandwidth of a typical burst of interference were not much less than the duration and bandwidth over which the profile was folded, as there would then be few independent instances of zapping within each profile, and so the binomial distribution of zapping might not resemble a Gaussian shape. Similarly, it could happen if the time and frequency resolutions of zapping were not much less than the folding duration and bandwidth. As a rough guide, fewer than $100$ independent instances of zapping within a profile may be too few when $q=0.1$. With or without zapping, it is worth noting that profile noise may be substantially skewed and non-Gaussian if few (less than about $50$) amplitude samples contribute to the power in each profile bin.

\section{Interference removal from pulsar observations}
\label{pulsar}

Spectral kurtosis has been employed successfully by the LEAP project, which makes astronomical observations of pulsars using up to five radio telescopes simultaneously \citep{bjk+16,sbj+17}. The aim of the project is to measure the times of arrival (TOAs) of pulses with sufficient accuracy to detect variation that is characteristic of the influence of gravitational waves, thereby measuring the strength of a background of low-frequency waves that is believed to permeate the Solar System from distant sources such as binary supermassive black holes \citep{hd83}. The signal from each telescope is converted to the baseband frequency range and sampled at the Nyquist rate to enable coherent summation \citep[pp. 117-120]{lk05}, allowing spectral kurtosis to be used effectively alongside a simpler method that zaps portions of the signal whose power deviates greatly from an expected value or from the power of neighbouring portions \citep[\S4.5]{bjk+16}. Each telescope's signal is recorded digitally using $8$ sampling bits, calibrated for polarization accuracy and then zapped if necessary, before the stored signals are summed with their amplitudes calibrated to maximise the SNR of the observation. Pulse profiles, showing the average radio emission from a pulsar as it rotates, can then be produced and timed.

\begin{figure}
\includegraphics[width=0.475\textwidth,clip=true,trim=0cm 0cm 0cm 0cm]{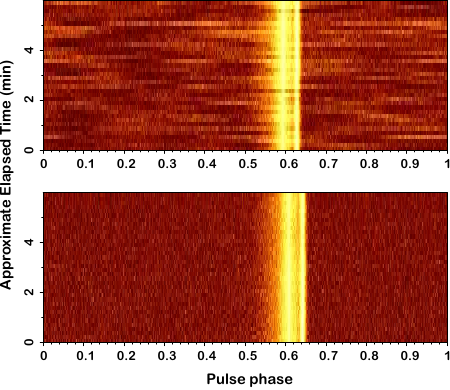}
\caption{The pulse profile of \mbox{PSR J1022$+$1001}, with brightness representing power, during six minutes of a LEAP observation with a bandwidth of $128\,\mathrm{MHz}$, both without (top) and with (bottom) interference removal using spectral kurtosis. The observation is a coherent summation of signals from the Jodrell Bank, Effelsberg, Nan\c{c}ay and Westerbork radio telescopes, with spectral kurtosis applied to Nan\c{c}ay using $M=1000$, $N=2$ and $\eta=3$, and rejected data replaced by artificial Gaussian noise. This high-resolution zapping makes the observation usable while sacrificing only a small fraction of data. Reproduced from \citet[Fig.~6]{bjk+16}.}
\label{fig:profile_1022}
\end{figure}

Fig.~\ref{fig:profile_1022}, reproduced from \citet{bjk+16}, shows the improvement in the pulse profile of \mbox{PSR J1022$+$1001} achieved by zapping a signal from the Nan\c{c}ay radio telescope using spectral kurtosis and replacing the rejected data with artificial Gaussian noise. The six-minute segment of this LEAP observation used four telescopes and covered a frequency range of $1332$--$1460\,\mathrm{MHz}$. Each measurement of the estimator used 1000 power values averaged over two complex polarization channels ($M=1000$ and $N=2$), giving zapping resolutions of $6.25\,\mathrm{ms}$ and $0.16\,\mathrm{MHz}$. The estimator thresholds were set using $\eta=3$, meaning that $0.27\ \mathrm{per\ cent}$ of good data from Nan\c{c}ay were zapped. Through the application of spectral kurtosis, little data were lost and an observation that was riddled with interference became suitable for high-precision pulsar timing.

\begin{figure}
\includegraphics[width=0.495\textwidth,clip=true,trim=0.9cm 0cm 0cm 1.96cm]{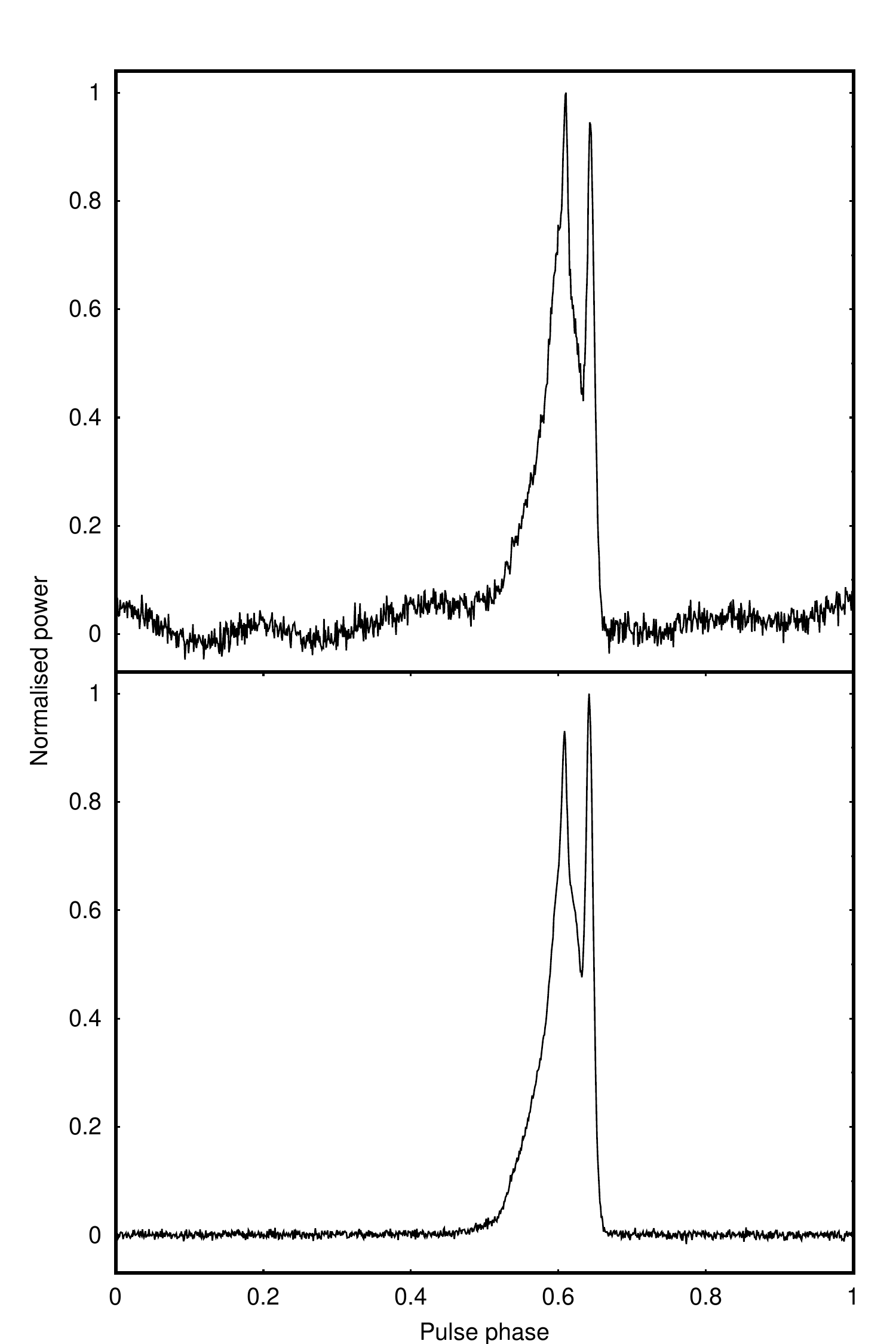}
\caption{The pulse profile of \mbox{PSR J1022$+$1001}, folded over 30 minutes of a LEAP observation with a bandwidth of $96\,\mathrm{MHz}$, both without (top) and with (bottom) interference removal using spectral kurtosis. The top panel shows the contribution from the Nan\c{c}ay radio telescope only, while the bottom panel shows the coherent summation of signals from the Jodrell Bank, Effelsberg, Nan\c{c}ay and Westerbork radio telescopes, with spectral kurtosis applied to Nan\c{c}ay using $M=1000$, $N=2$ and $\eta=3$, and rejected data replaced by artificial Gaussian noise. Zapping restores the profile shape and flattens the power baseline, allowing the pulse arrival time to be measured with a high estimated accuracy of $0.25\,\mathrm{\mu s}$. This compares to $1.35\,\mathrm{\mu s}$ for Nan\c{c}ay alone and without zapping.}
\label{fig:profile_1022_folded}
\end{figure}

\begin{figure}
\includegraphics[width=0.576\textwidth,clip=true,trim=3.06cm 2.45cm 0cm 3.2cm]{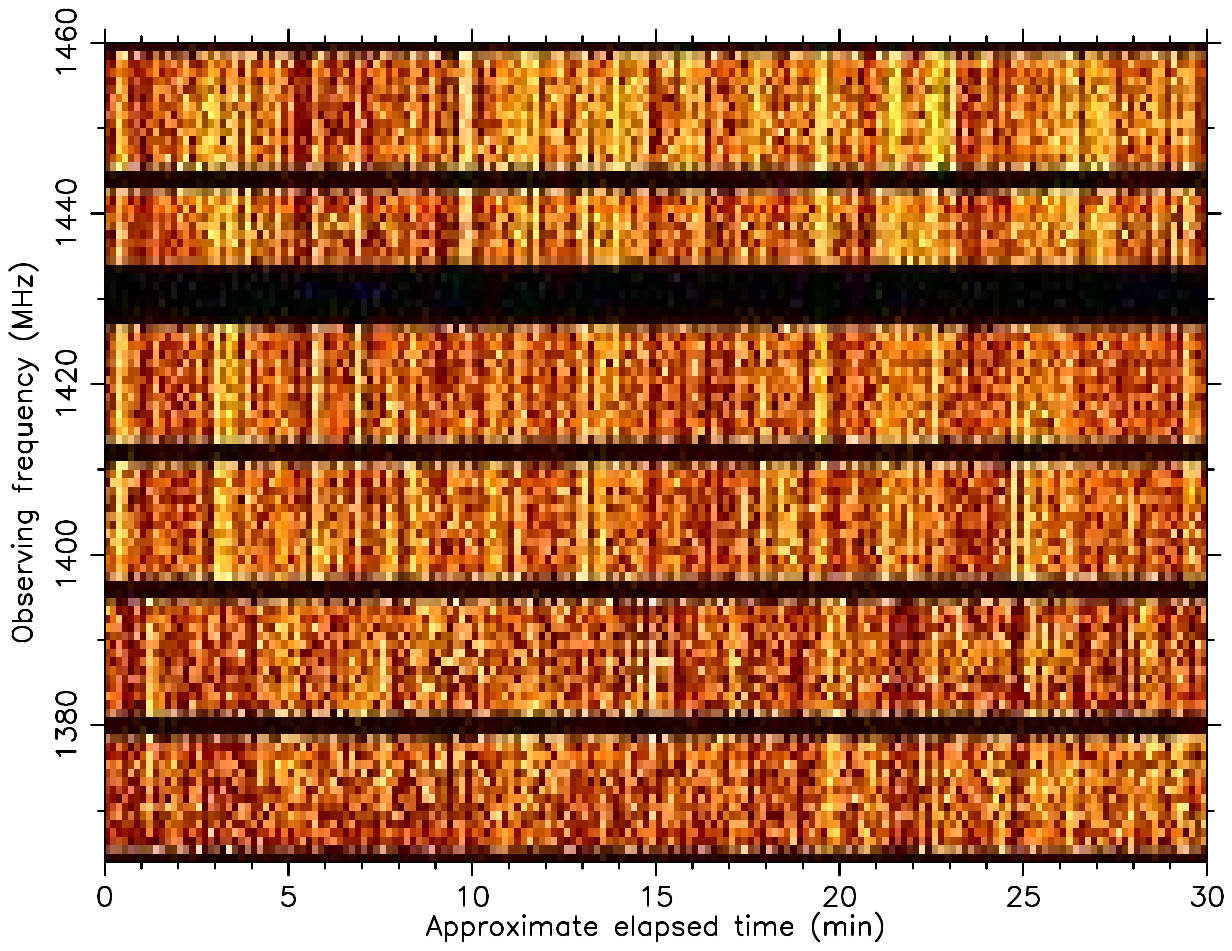}
\caption{The dynamic spectrum of the Nan\c{c}ay observation shown in Fig.~\ref{fig:profile_1022_folded}, with brightness representing power, before interference removal. Spectral kurtosis is applied before the pulsar observation is folded, and can zap the persistent broadband interference seen here.}
\label{fig:rfi_1022}
\end{figure}

Fig.~\ref{fig:profile_1022_folded} shows the improvement in the pulse profile shape of \mbox{PSR J1022$+$1001} achieved by zapping a signal from the Nan\c{c}ay radio telescope using the same parameters and resolutions as those above, while \mbox{Fig.~\ref{fig:rfi_1022}} reveals the persistent broadband interference that was removed. This 30-minute LEAP observation took place on 2013 July 27, used four telescopes and covered a frequency range of $1364$--$1460\,\mathrm{MHz}$. Following the application of spectral kurtosis, the \textsc{psrchive} tool \textsc{pat} was used to align the LEAP profile with a template profile of high S/N using the Fourier phase gradient between them \citep{tay92}. This gave an estimated uncertainty of only $0.25\,\mathrm{\mu s}$ in the pulse arrival time associated with the zapped tied-array profile, compared to $1.35\,\mathrm{\mu s}$ for the non-zapped Nan\c{c}ay profile alone.

\begin{figure}
\includegraphics[width=0.495\textwidth,clip=true,trim=0.9cm 0cm 0cm 1.96cm]{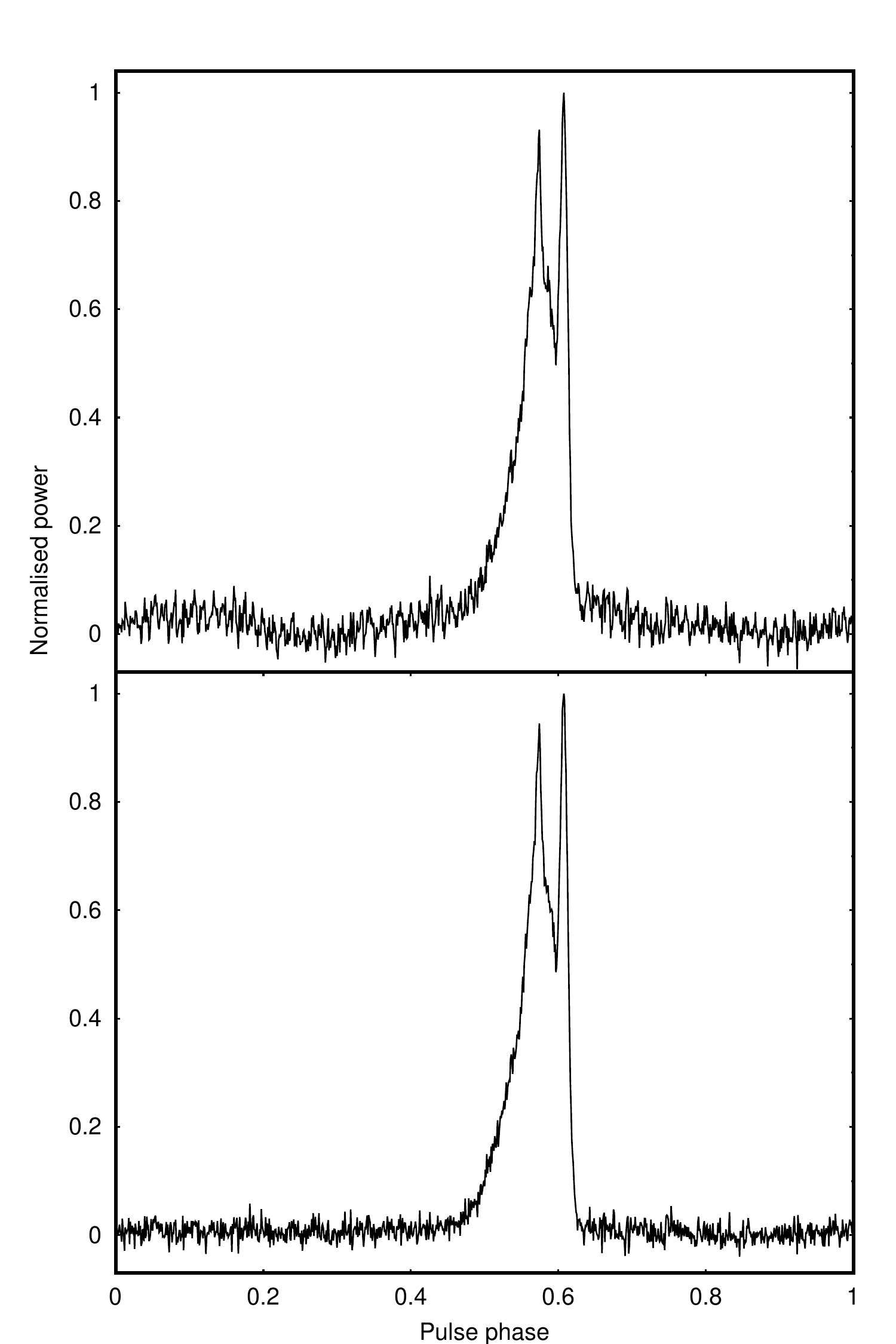}
\caption{The pulse profile of \mbox{PSR J1022$+$1001}, folded over 60 minutes of a LEAP observation with a bandwidth of $128\,\mathrm{MHz}$, both without (top) and with (bottom) interference removal using spectral kurtosis. Both panels show the contribution from the Nan\c{c}ay radio telescope only, with spectral kurtosis applied using $M=1000$, $N=2$ and $\eta=3$, and rejected data replaced by artificial Gaussian noise. While zapping of the folded observation is ineffective at removing this interference, spectral kurtosis improves the estimated accuracy of the pulse arrival time measured at Nan\c{c}ay from $1.22$ to $0.70\,\mathrm{\mu s}$.}
\label{fig:profile_1022_folded_new}
\end{figure}

\begin{figure}
\includegraphics[width=0.475\textwidth,clip=true,trim=0cm 0cm 0cm 0cm]{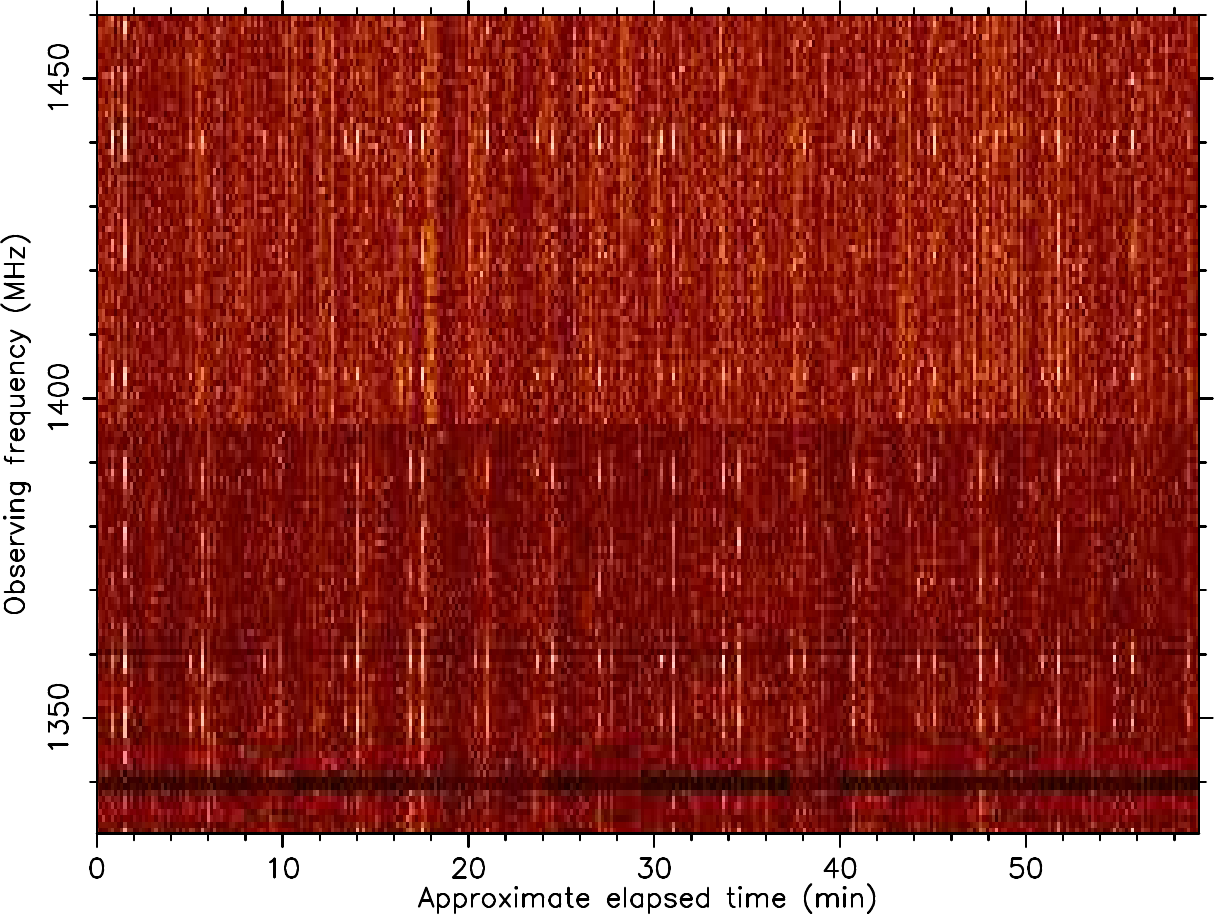}
\caption{The dynamic spectrum of the Nan\c{c}ay observation shown in Fig.~\ref{fig:profile_1022_folded_new}, with brightness representing power, before interference removal. The persistent broadband interference seen here cannot be separated from the pulsar signal using a folded observation, but can be zapped by spectral kurtosis using fine time and frequency resolutions simultaneously.}
\label{fig:rfi_1022_new}
\end{figure}

\begin{figure}
\includegraphics[width=0.495\textwidth,clip=true,trim=3.55cm 0.4cm 5.5cm 1.55cm]{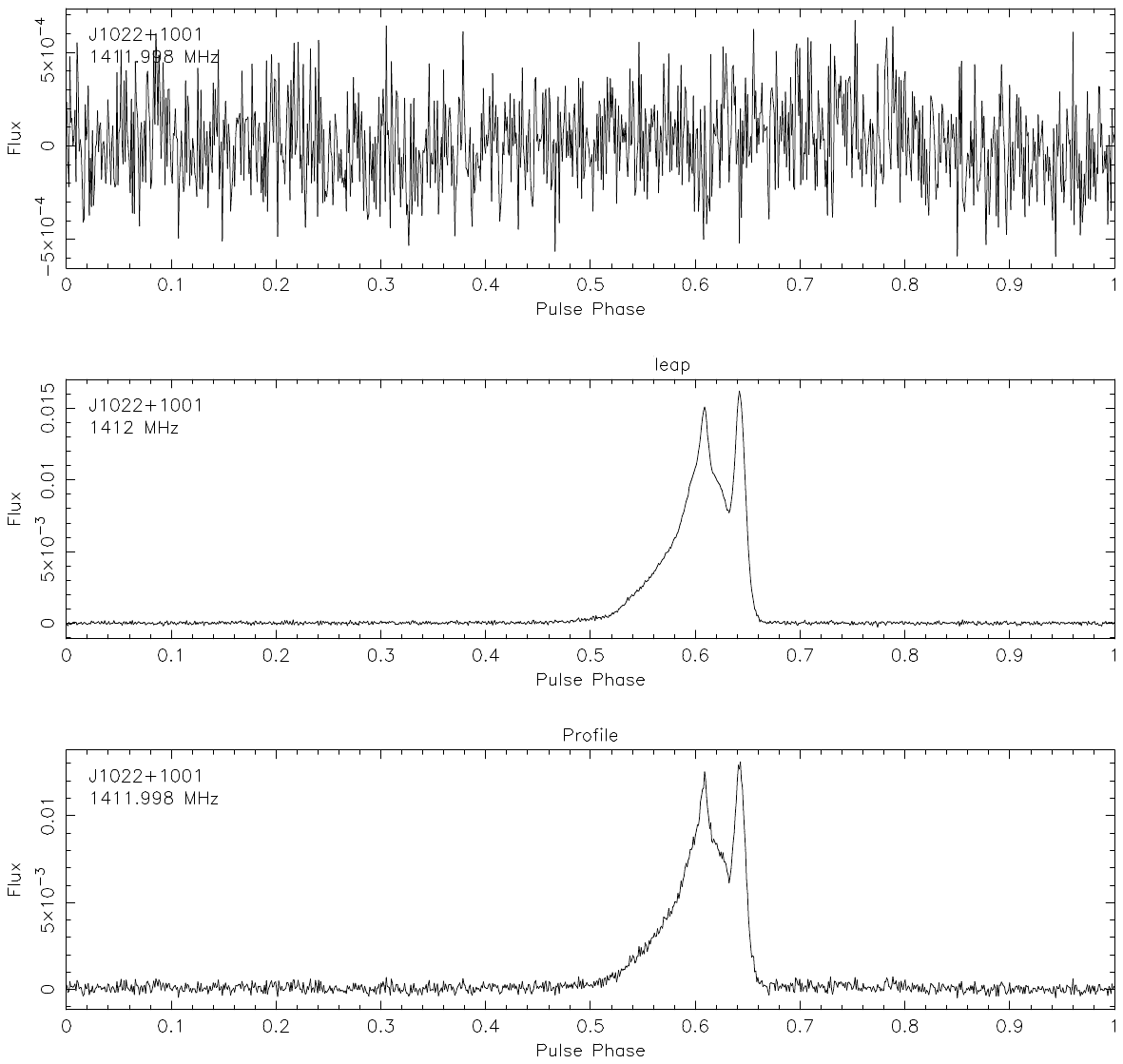}
\caption{The pulse profile of \mbox{PSR J1022$+$1001}, with a residual profile (top) showing the difference between the four-telescope LEAP observation from Fig.~\ref{fig:profile_1022_folded} (middle) and the Nan\c{c}ay observation from Fig.~\ref{fig:profile_1022_folded_new} (bottom) after interference removal using spectral kurtosis. The Nan\c{c}ay observation has been reduced to the same frequency range as the earlier LEAP observation, and its profile has been scaled and shifted when producing the residual profile. Interference removal from the Nan\c{c}ay telescope does not appear to have changed the pulse profile shape, judging by the lack of structure in the residuals between these two observations that were made eight years apart.}
\label{fig:profile_1022_residual_leap}
\end{figure}

A further LEAP observation of \mbox{PSR J1022$+$1001} demonstrates that spectral kurtosis recovers the unique profile shape that is critical to high-precision pulsar timing. Fig.~\ref{fig:profile_1022_folded_new} shows the profile of \mbox{PSR J1022$+$1001} produced from a 60-minute observation made on 2021 May 15 by the Nan\c{c}ay radio telescope over a frequency range of $1332$--$1460\,\mathrm{MHz}$, before and after zapping using the same parameters and resolutions as those above. Fig.~\ref{fig:rfi_1022_new} shows that the interference in the observation was persistent and broadband, and zapping of the folded observation using the \textsc{psrchive} tool \textsc{paz} had little effect. Spectral kurtosis zapped around $0.6\ \mathrm{per\ cent}$ of the data, and reduced the estimated timing uncertainty of the Nan\c{c}ay profile from $1.22$ to $0.70\,\mathrm{\mu s}$ according to the \textsc{pat} tool. Fig.~\ref{fig:profile_1022_residual_leap} shows the residual profile produced by subtracting this Nan\c{c}ay observation from the four-telescope LEAP observation shown in Fig.~\ref{fig:profile_1022_folded}, again using \textsc{pat}, with the lack of residual structure indicating that spectral kurtosis did not alter the shape of the pulse profile.

\section{Conclusions}
\label{conc}

This paper provides a recipe for the implementation of the spectral kurtosis method from start to finish, allowing signal interference to be zapped from real or complex time series data stored as either amplitudes or power (Section~\ref{method}). The frequentist nature of spectral kurtosis makes it effective without prior knowledge of the interference that will be encountered, so it is widely applicable rather than being ideal in specific situations (Section~\ref{advs}). We have shown its success in enabling an accurate radio-frequency array observation of a pulsar in the presence of interference local to one telescope, allowing signals from multiple widely spaced telescopes to be combined with only a very small loss of usable information and without any apparent detriment to the shape of the pulse profile (Section~\ref{pulsar}). The preservation of the unique profile signature of each pulsar is crucial for precise timing of its rotation, and the timing information from the cleaned observations is being used in a long-term project to detect gravitational waves.

When zapping data that contain a rapidly varying signal such as pulsar emission, it is important that the estimator does not recognize the signal amplitudes as non-Gaussian, as the spectral kurtosis procedure would then remove the information of interest. Observers should therefore ensure that the time and frequency resolutions of an observation are too fine to allow single pulses to be detected (Section~\ref{advs}). In order to maintain a Gaussian noise distribution, the time and frequency resolutions of zapping should be much less than the duration and bandwidth of an observation or folded pulse profile, so that there are many independent opportunities for zapping within the observation (Section~\ref{effects}).

The quality of an observation made using an array of telescopes in independent interference environments is improved by compensating for zapped data, regardless of the zapping technique used, and the methods of compensation are applicable to any widely spaced array \mbox{(Section~\ref{effects})}. The highest SNR is usually obtained by mean equalization: equalizing the mean of the summed power so that its baseline level remains constant over time. Mean equalization is the most appropriate method for an incoherently summed signal. However, if the signal is coherently summed and its amplitudes are stored so that its time and frequency resolutions can be adjusted later, it is better to apply variance equalization by scaling: equalizing the variance of the summed amplitudes over time, which also results in a constant power baseline and avoids unwanted artefacts if the amplitudes are transformed to different time and frequency resolutions. The SNR after variance equalization by scaling is usually slightly less than the SNR after mean equalization, but the difference is small when there is either a small-to-moderate amount of zapping or a large number of telescopes. Variance equalization by scaling is the most appropriate method when the signal may be summed coherently or incoherently in different parts of an observation. The alternative method of variance equalization by addition of artificial noise may be needed to allow signals from multiple telescopes to be synchronised, and can then be replaced with another method to improve SNR.

\section*{Acknowledgements}

The European Pulsar Timing Array (EPTA) is a collaboration of European institutes working towards the direct detection of low-frequency gravitational waves, for which it has implemented the Large European Array for Pulsars (LEAP). The authors acknowledge the support of colleagues in the EPTA; MP expresses gratitude for the patience of co-authors and family while the writing of this paper was completed, and dedicates the paper to his father. The work reported here has been funded by the European Research Council Advanced Grant ``LEAP'', Grant Agreement ID 227947 (Principal Investigator: M. Kramer). KL and MK are supported by the European Research Council Synergy Grant ``BlackHoleCam'', Grant Agreement ID 610058 (Principal Investigator: M. Kramer). KJL acknowledges support from the National Basic Research Program of China, 973 Program, 2015CB857101 and NSFC 11373011. The Effelsberg Radio Telescope is operated by the Max-Planck-Institut f\"{u}r Radioastronomie in Germany. The Westerbork Synthesis Radio Telescope is operated by the Netherlands Institute for Radio Astronomy (ASTRON) with support from the Netherlands Organisation for Scientific Research (NWO). The Nan\c{c}ay Radio Observatory is operated by the Paris Observatory and associated with the Centre National de la Recherche Scientifique in France. Pulsar research at the Jodrell Bank Centre for Astrophysics, and the observations using the Lovell Telescope, are supported by a consolidated grant from the Science and Technology Facilities Council (STFC) in the UK. The Sardinia Radio Telescope is operated by the Istituto Nazionale di Astrofisica (INAF) in Italy, and was undergoing its astronomical validation phase when the observations used in this paper were made.

\section*{Data availability}

The data and software programmes underlying this article will be shared on reasonable request to the corresponding author.
\bibliographystyle{mnras}
\bibliography{journals,psrrefs,modrefs,crossrefs}

% Typesetting comment
\bsp
\label{lastpage}
\end{document}